\documentclass{article}[11pt]

\usepackage{amsmath,amssymb,amsfonts}
\usepackage{physics}
\usepackage{dsfont}
\usepackage{graphicx}
\usepackage{float}
\usepackage{tikz}

\usetikzlibrary{arrows,shapes,trees}

\newcommand{\ra}{\rightarrow}

\newcommand{\End}{{\rm End}}

\newcommand{\CC}{{\mathbb C}}
\newcommand{\ZZ}{{\mathbb Z}}

\newcommand{\cP}{{\mathcal P}}

\newcommand{\cA}{\mathcal A}

\newcommand{\cG}{{\mathcal G}}

\newcommand{\cT}{{\mathcal T}}
\newcommand{\cF}{{\mathcal F}}
\newcommand{\cQ}{{\mathcal Q}}

\newcommand{\Cl}{{C\ell}}

\newcommand{\fotimes}{\mathbin{\widehat\otimes}}
\newcommand{\Arf}{{\rm Arf}}

\usepackage[left=1in,right=1in,top=1in,bottom=1in]{geometry}

\usepackage{hyperref}

\usepackage{subcaption}

\usetikzlibrary{positioning,calc}
\usetikzlibrary{decorations.markings}

\hypersetup{colorlinks=true,urlcolor=[rgb]{0,0,0.5},citecolor=[rgb]{0,0.5,0},linkcolor=[rgb]{0,0.5,0}}

\title{Spin Topological Field Theory and Fermionic Matrix Product States}
\author{Anton Kapustin, Alex Turzillo, Minyoung You \\
{\it California Institute of Technology, Pasadena, CA 91125}}

\begin{document}
\maketitle

\begin{abstract}
We study state-sum constructions of $G$-equivariant spin-TQFTs and their relationship to Matrix Product States. We show that in the Neveu-Schwarz, Ramond, and twisted sectors, the states of the theory are generalized Matrix Product States. We apply our results to revisit the classification of fermionic Short-Range-Entangled phases with a unitary symmetry $G$ and determine the group law on the set of such phases. Interesting subtleties appear when the total symmetry group is a nontrivial extension of $G$ by fermion parity.
\end{abstract}


\section{Introduction and Overview}

Recently the problem of classifying Short-Range-Entangled (SRE) phases of matter has attracted considerable attention. A powerful approach for 1d systems is the Matrix Product State representation of ground states (see \cite{MPSreview} for a review). For bosonic systems with a symmetry $G$, this leads to a classification of SRE phases in terms of group cohomology of $G$ \cite{ChenGuWenone,fidkit}. Fermionic systems in 1d are related to bosonic systems with a $\ZZ_2$ symmetry via the Jordan-Wigner transformation. This enables one to classify 1d fermionic  SRE phases of matter as well \cite{fidkit,ChenGuWentwo}.

There is a conjectural  classification of SRE phases in all dimensions \cite{kapustincobord,KTTW} (see also \cite{Freed}) based on entirely different ideas. In the case of bosonic (resp. fermionic) SRE phases phases with an internal finite symmetry $G$ in $d$ spatial dimensions, the conjecture says that they are classified by the torsion part of the $(d+1)$-dimensional oriented cobordism (resp. spin-cobordism) of $BG$ with $U(1)$ coefficients. Here $BG$ is a certain infinite-dimensional topological space  known as the classifying space of $G$. This conjecture is partially explained by the recently proved mathematical theorem \cite{FreedHopkins} which states that oriented (resp. spin) $(d+1)$-dimensional cobordism groups classify unitary invertible oriented (resp. spin) Topological Quantum Field Theories in $d+1$ space-time dimensions. This is only a partial explanation, because the relation between SRE phases and TQFTs remains conjectural. In 1d, one could hope for a more direct connection between the cobordism/TQFT data and the MPS data.

For bosonic SRE phases in 1d the connection between the MPS approach and the cobordism/TQFT approach has been recently clarified \cite{ShiozakiRyu,KTY}. In particular, it has been shown in \cite{KTY}  that an MPS representation of ground states naturally arises from an annulus diagram in a TQFT. The goal of this paper is  to extend this observation to spin-TQFTs and the associated fermionic MPS.

Let us describe the structure of the paper and the main results. In section 2, we review the state-sum construction of spin-TQFTs in two space-time dimensions from $\ZZ_2$-graded algebras following \cite{NR,GK}. We also show that stacking fermionic systems together corresponds to taking the supertensor product of the corresponding algebras. This gives a very clean and simple derivation of the spin-statistics relation in the topological case. In section 3, we evaluate the annulus diagram and show that it gives rise to a generalized MPS both in the Neveu-Schwarz and the Ramond sector. In section 4 we work out the commuting projector Hamiltonian starting from the TQFT data describing an invertible spin-TQFT. We show that for a nontrivial spin-TQFT the resulting Hamiltonian describes the Majorana chain \cite{fidkit}. In section 5, we discuss $G$-equivariant spin-TQFT and $G$-equivariant fermionic MPS. We show that fermionic SRE phases with a symmetry $G$ times the fermion parity are in 1-1 correspondence with invertible $G$-equivariant spin-TQFTs, and that the TQFT data give rise to fermionic $G$-equivariant MPS. We also discuss the case when the symmetry is a nontrivial extension $\cG$ of $G$ by fermion parity, which is related to $\cG$-Spin TQFTs. In all cases we determine the group law on the set of fermionic SRE phases. Finally, we discuss in some detail fermionic SRE phases with symmetry $\ZZ_2$. 

While this paper was in preparation, a preprint \cite{FMPS} appeared on the arXiv which also describes $G$-equivariant fermionic MPS.

The research of A.K. was supported by the U.S. Department of Energy, Office of Science, Office of High Energy Physics, under Award Number DE-SC0011632, and by the Simons Investigator Award.

\section{Spin-TQFTs}\label{spintqft}


\subsection{$\ZZ_2$-graded semi-simple algebras}

The algebraic input for the fermionic state-sum construction is a $\ZZ_2$-graded semisimple Frobenius algebra $A$ \cite{NR,GK}.\footnote{While it is possible to relax the semi-simplicity condition \cite{NR}, here we are interested in unitary TQFTs, and for such TQFTs one may assume that $A$ is semi-simple \cite{KTY}.} A Frobenius algebra is a finite-dimensional algebra over $\CC$ with a non-degenerate symmetric scalar product $\eta:A\otimes A\ra\CC$ satisfying $\eta(a,bc)=\eta(ab,c)$ for all $a,b,c\in A$.  A $\ZZ_2$-grading on $A$ is a decomposition $A=A_+\oplus A_-$ such that 
\begin{equation}
A_+\cdot A_+\subset A_+,\quad A_-\cdot A_-\subset A_+,\quad A_-\cdot A_+\subset A_-,\quad A_+\cdot A_-\subset A_-.
\end{equation}
Equivalently, a $\ZZ_2$-grading is an operator $\cF:A\ra A$ such that $\cF^2=1$ and $\cF(a)\cdot \cF(b)=\cF(a\cdot b)$. The operator $\cF$ is called fermion parity and is  traditionally denoted $(-1)^F$. We also assume that the scalar product $\eta$ is $\cF$-invariant:
\begin{equation}\eta(\cF(a),\cF(b))=\eta(a,b).
\end{equation}


Note that $\cF$ defines an action of $\ZZ_2$ on $A$ which makes $A$ into a $\ZZ_2$-equivariant algebra. This observation is the root cause of the bosonization phenomenon: there is a 1-1 map between 1+1d phases of bosons with $\ZZ_2$ symmetry and 1+1d phases of fermions. For now, we use this fact to describe the classification of $\ZZ_2$-graded simple algebras. Namely, since the only proper subgroup of $\ZZ_2$ is the trivial one, and $H^2(\ZZ_2,U(1))=0$, a simple $\ZZ_2$-graded algebra is isomorphic either to $\End(V)$ for some $\ZZ_2$-graded vector space $V=V_+\oplus V_-$, or to $\Cl(1)\otimes\End(V)$ for some purely even vector space $V=V_+$ \cite{KTY}. Here $\Cl(1)$ denotes the Clifford algebra with one generator, i.e. an algebra with an odd generator $\Gamma$ satisfying $\Gamma^2=1$.

As explained in \cite{KTY}, the bosonic phase depends only on the Morita-equivalence class of $A$.
The choice of $V$ does not affect the Morita-equivalence class of the algebra, so there are only two Morita equivalence classes of $\ZZ_2$-graded algebras: the trivial one, corresponding to the algebra $\CC$, and the nontrivial one, corresponding to the algebra $\Cl(1)$. In the bosonic case, the former one corresponds to the trivial gapped phase with a $\ZZ_2$ symmetry, while the latter one corresponds to the phase with a spontaneously broken $\ZZ_2$. 

The fermionic interpretation is different. As briefly mentioned in \cite{GK} and discussed in more detail below, the algebra $\Cl(1)$ describes a gapped  fermionic phase which is equivalent to the nontrivial  Majorana chain. This is in accord with the intuition that fermion parity cannot be spontaneously broken.


\subsection{Spin structures}\label{spinstructsection}

A spin structure on an oriented manifold enables one to define a spin bundle. For a 1d manifold $X$, a spin bundle is a real line bundle $L$ plus an isomorphism $L\otimes L\ra TX$. Thus a spin bundle is a square root of the tangent bundle. Since $TX$ is trivial, such $L$ are classified by elements of $H^1(X,\ZZ_2)$. Since $H^1(S^1,\ZZ_2)=\ZZ_2$, there are two possible spin structures on a circle, called the R (Ramond) an NS (Neveu-Schwarz) spin structures in the string theory literature. The R structure corresponds to a trivial $L$, while NS structure corresponds to the ``M\"obius band'' $L$. In other words, if we give $L$ a metric and compute the holonomy of the unique connection compatible with it along $S^1$, we get $1$ for the R case, and $-1$ for the NS case. 

For an oriented 2d manifold $\Sigma$, we can regard $T\Sigma$ as a complex line bundle, and then a spin bundle on $\Sigma$ is a complex line bundle $S$ equipped with an isomorphism $S\otimes S\otimes S\ra T\Sigma$. One can show that such an $S$ always exists. Clearly, if $S$ and $S'$ are two spinor bundles, they differ by a line bundle which squares to a trivial line bundle on $\Sigma$. The latter are classified by elements of $H^1(\Sigma,\ZZ_2)$. Thus there are as many spin structures as there are elements of $H^1(\Sigma,\ZZ_2)$. But in general there is no natural way to identify elements of $H^1(\Sigma,\ZZ_2)$ with spin structures.\footnote{The case of a torus is an exception, since then $T\Sigma$ is trivial. This is why one can talk about periodic and anti-periodic spin structures on a torus.}

It is easy to see that a spin structure $s$ on an oriented 2d manifold $\Sigma$ induces a spin structure on any oriented 1d manifold $\gamma$ embedded into $\Sigma$. Define $\sigma_s(\gamma)=+1$ if the induced structure is of the NS type and $\sigma_s(\gamma)=-1$ if the induced structure is of the R type. That is, $\sigma_s(\gamma)$ is the negative of the holonomy of the connection corresponding to the induced spin structure. It is easy to show that $\sigma_s(\gamma)$ depends only on the homology class of $\gamma$ and thus defines a function $\sigma_s: H_1(\Sigma,\ZZ_2)\ra\ZZ_2$. With more work, one can show that this function satisfies
\begin{equation}\label{quad}
\sigma_s([\gamma]+[\gamma'])=\sigma_s([\gamma])\sigma_s([\gamma'])(-1)^{\langle [\gamma],[\gamma']\rangle}.
\end{equation}
That is, it is a quadratic $\ZZ_2$-valued function on $H_1(\Sigma,\ZZ_2)$ whose corresponding bilinear form is the intersection pairing on $H_1(\Sigma,\ZZ_2)$. In fact, it is a theorem of Atiyah \cite{Atiyah} that for a closed $\Sigma$ the spin structure is determined by such a quadratic function, and that any such quadratic function determines a spin structure. Note that the ratio of two such quadratic functions is a linear function on $H_1(\Sigma,\ZZ_2)$, or equivalently an element of $H^1(\Sigma,\ZZ_2)$. Thus we recover the result that two spin structures differ by an element of $H^1(\Sigma,\ZZ_2)$. 

We record for future use another property of the function $\sigma_s$:
\begin{equation}\label{sigmasvariation}
\sigma_{s+a}([\gamma])=(-1)^{\int_\gamma a}\sigma_s([\gamma]),
\end{equation}
where $a$ is an arbitrary element of $H^1(\Sigma,\ZZ_2)$. Thus $\sigma_s([\gamma])$ is an affine-linear function of $s$ and a quadratic function of $[\gamma]$. 

We will also need a version of this result for the case when $\Sigma$ has a nonempty boundary. As in the case of equivariant TQFT, it is convenient to choose, along with a spin structure $s$, a point on every connected component of $\partial\Sigma$ and a normalized basis vector for the real spin bundle $L$ at this point. This simplifies the gluing of spin manifolds. We will denote by $\partial_0\Sigma$ the set of all marked points, and will call a spin structure on $\Sigma$ together with a trivialization of $L$ at $\partial_0\Sigma$ a spin structure on the pair $(\Sigma,\partial_0\Sigma)$. The group $H^1(\Sigma,\partial_0\Sigma;\ZZ_2)$ acts freely and transitively on the set of spin structures on $(\Sigma,\partial_0\Sigma)$. Despite this, there is no canonical way to identify spin structures with elements of $H^1(\Sigma,\partial_0\Sigma;\ZZ_2)$. To get an algebraic description of spin structures, one can proceed as follows \cite{Segal}. First, note that $H_1(\Sigma,\partial_0\Sigma;\ZZ_2)$ can be identified with $H_1(\Sigma_*,\ZZ_2)$, where $\Sigma_*$ is a closed oriented 2d manifold obtained by gluing a sphere with holes onto $\Sigma$. This identification depends on the choice of a cyclic order of the set of boundary circles of $\Sigma$. Thus the intersection form on $H_1(\Sigma_*,\ZZ_2)$ induces a non-degenerate symmetric bilinear form on $H_1(\Sigma,\partial_0\Sigma;\ZZ_2)$.  There is also an identification of the set of spin structures on $(\Sigma,\partial_0\Sigma)$ and the set of of spin structures on $\Sigma^*$ \cite{Segal}. Thus the set of spin structures on $(\Sigma,\partial_0\Sigma)$ can be identified with the set of $\ZZ_2$-valued quadratic functions on $H_1(\Sigma,\partial_0\Sigma;\ZZ_2)$ refining the intersection form. This identification still depends on a choice of a cyclic order on the set of boundary circles of $\Sigma$. One can determine which spin structure is induced on any particular connected component of $\partial\Sigma$ by evaluating  this quadratic function on the closed curve wrapping that component.

\subsection{State-sum construction of the spin-dependent  partition function}

To define the partition function of a spin-TQFT on a closed oriented 2-manifold $\Sigma$ with a spin structure, 
we choose a skeleton of $\Sigma$, i.e. a trivalent graph $\Gamma$ on $\Sigma$ whose complement is homeomorphic to a disjoint union of disks. Equivalently, one may think of $\Gamma$ as the Poincar\'{e} dual of a triangulation $\cT$ of $\Sigma$.\footnote{One can formulate the construction either in terms of triangulations or in terms of skeletons, but the latter approach gives a bit more flexibility when we allow $\Sigma$ to have a nonempty boundary.} For every vertex $v\in\Gamma$, let $\Gamma(v)$ denote the edges containing $v$. Orientation of $\Sigma$ gives rise to a cyclic order on $\Gamma(v)$ for all $v$. This is sufficient to produce the partition function of a bosonic TQFT based on the algebra $A$, but in order to construct the fermionic partition function, we need to choose an actual order on $\Gamma(v)$. We can do it by picking one special edge $e_0(v)\in\Gamma(v)$ for every $v$. We also choose an orientation for each edge of $\Gamma$. (In Ref. \cite{GK} both an orientation of edges and a choice of $e_0(v)$ arose from a branching structure on $\cT$, but here we follow Ref. \cite{NR} and choose them independently.) These choices are called a \emph{marking} of $\Gamma$.


We also need to describe a choice of spin structure on $\Sigma$. This is a cellular 1-cochain $s$ valued in $\ZZ_2$ (i.e. an assignment of elements of $\ZZ_2$ to edges of $\Gamma$) with coboundary a certain 2-cocycle $w_2$ whose cohomology class is the second Stiefel-Whitney class $[w_2](\Sigma)$. Following Ref. \cite{NR}, we write the constraint $\delta s=w_2$ as
\begin{equation}\label{w2}
(\delta s)(f)=1+K+D\ {\rm mod}\ 2.
\end{equation}
where $f$ is a particular cell in $\Sigma\backslash \Gamma$, $K$ is the number of clockwise oriented edges in $\partial f$, 
and $D$ is the number of vertices $v$ for which the counterclockwise-oriented curve homologous to $\partial f$ in $\Gamma$ enters $v$ through $e_0(v)$. Two solutions $s,s'$ of this constraint are regarded equivalent, $s\sim s'$, if $s-s'=\delta t$ for some 0-cochain $t$. Two solutions $s,s'$ define isomorphic spin structures on $\Sigma$ if and only if $s\sim s'$ \cite{NR,GK}. Thus we recover the fact that the number of distinct spin structures on $\Sigma$ is equal to $|H^1(\Sigma,\ZZ_2)|$.


One can give an explicit description of the holonomy function $\sigma_s(\gamma)$ corresponding to the 1-cochain $s$. Regard a closed curve $\gamma$ embedded into the graph $\Gamma$ as a $1$-cycle. Then $\sigma_s(\gamma)$ is given in terms of the signs $s$ and the marking of $\Gamma$ along $\gamma$ by eq. $(3.45)$ of Ref. \cite{NR}. This expression simplifies greatly in the important case of when $\gamma$ is a counterclockwise-oriented curve bounding a single cell in $\Sigma\backslash\Gamma$. Here $\sigma_s(\gamma)=-(-1)^{s(\gamma)}$; that is, $-1$ for each edge of $\gamma$ oriented clockwise, times $-1$ for each vertex $v$ such that $\gamma$ enters $v$ through $e_0(v)$. One can show that this function depends only on the homology class of $\gamma$ and is a quadratic refinement of the intersection form.

Choose a basis $e_i$ in $A$ whose elements are eigenvectors of $\cF$. Let $\eta_{ij}=\eta(e_i,e_j)$. Since $\eta$ is non-degenerate, it has an inverse $\eta^{ij}$. Let $C^i{}_{jk}$ denote the structure constants of $A$. Define $C_{ijk}=\eta_{il}C^l{}_{jk}$. It can be shown that the tensor $C_{ijk}$ is cyclically symmetric \cite{KTY}. Denote by $(-1)^{\beta_i}$ the eigenvalue of $\cF$ corresponding to $e_i$.



Now we can explain the recipe for computing the partition function for a surface $\Sigma$ with a marked skeleton $\Gamma$ and a spin structure $s$. Each edge of $\Gamma$ is colored with a pair of basis vectors $e_i\in A$, and we have a factor of $C_{ijk}$ for each vertex and $\eta^{ij}$ for each edge. Since $A$ is $\ZZ_2$-graded, $\eta^{ij}$ vanishes unless $\beta_i = \beta_j$, and $C_{ijk}$ vanishes unless $\beta_i + \beta_j + \beta_k = 0$. Hence the function $\beta: e_i\mapsto \beta_i$ on the set of edges of $\Gamma$ defines a {\rm mod}-2 1-cycle on $\Sigma$. The contribution of a particular coloring of $\Gamma$ is the product of all $C_{ijk}$ and $\eta^{ij}$, the spin-dependent sign factor
\begin{equation}\label{spinsign}
(-1)^{s(\beta)}=(-1)^{\sum_e s(e)\beta(e)},
\end{equation}
and the Koszul sign $\sigma_0(\beta)$. The partition function is obtained by summing over all colorings. Note that
\begin{equation}\label{fermstatesum}
Z_\text{ferm}(A,\eta)=\sum_{\beta} Z_\text{bose}(A,\beta) \sigma_s(\beta),
\end{equation}
where $Z_\text{bose}(A,\beta)$ is the sum over all colorings with a fixed 1-cycle $\beta$. Using the isomorphism $H_1(\Sigma,\ZZ_2)\simeq H^1(\Sigma,\ZZ_2)$, one can interpret $\beta$ as a $\ZZ_2$ gauge field on a dual triangulation and $Z_\text{bose}(A,\beta)$ as the partition function of a bosonic system with a global $\ZZ_2$ symmetry coupled to $\beta$. Equation (\ref{fermstatesum}) is a manifestation of the bosonization phenomenon. 

It remains to explain how the Koszul sign $\sigma_0(\beta)$ is evaluated. Consider a vertex whose edges are labeled by $i,j,k$ starting from the special edge and going counterclockwise. Assign to it an element $C_v= C_{ijk} e_i\otimes e_j\otimes e_k$ in $A\otimes A\otimes A$. Tensoring over vertices, we get an element $C_\Gamma$ of $A^{\otimes 3N}$, where $N$ is the number of vertices of $\Gamma$. Now consider an oriented edge of $\Gamma$ labeled by $i,j$. It corresponds to an ordered pair of factors in $C_\Gamma$. Permute the factors of $C_\Gamma$ until these two are next to each other and in order, keeping track of the fermionic signs\begin{equation}e_i\otimes e_j\mapsto (-1)^{\beta_i\beta_j}e_j\otimes e_i\end{equation}one incurs in the process, and then contract using the scalar product $\eta$. Continuing in this fashion, we are left with the product of all $C_{ijk}$ and $\eta^{ij}$ times a sign. This sign is the Koszul sign $\sigma_0(\beta)$. It is clear that it depends on the coloring of $\Gamma$ only through the 1-cycle $\beta$. Note that the elements $C_v$ as well as the pairs of factors for each edge are all even, so one does not need to order the set of vertices or the set of edges. One can also define $\sigma_0(\beta)$ as a Grassmann integral, as was  originally done in \cite{GuWen}. The product of the Koszul sign $\sigma_0(\beta)$ and the spin-dependent factor $(-1)^{s(\beta)}$ is nothing but the quadratic function $\sigma_s(\beta)$ \cite{GK}.

One can show \cite{NR,GK} that the partition function thus defined depends only on the spin surface $(\Sigma,s)$ and not the skeleton $\Gamma$, its marking, or the particular $1$-cochain representing $s$. Finally, it is clear that if $A$ is purely even, both the Koszul sign and the spin-dependent sign factor are trivial, and the partition function reduces to the bosonic partition function associated with $A$.

\subsection{Stacking and the supertensor product}

It is interesting to determine the behavior of the partition function under stacking systems together. Given a pair of fermionic systems encoded in a pair of $\ZZ_2$-graded Frobenius algebras $A_1,A_2$, stacking these systems together gives us a system with a partition function $Z_\text{ferm}(A_1,\eta) Z_\text{ferm}(A_2,\eta)$. It turns out that 
\begin{equation}\label{fermstacking}
Z_\text{ferm}(A_1,\eta) Z_\text{ferm}(A_2,\eta)=Z_\text{ferm}(A_1\fotimes A_2,\eta),
\end{equation}
where $\fotimes$ is the supertensor product of $\ZZ_2$-graded algebras. Let us recall what this means. The usual tensor product of algebras $A_1\otimes A_2$ obeys the multiplication rule
\begin{equation}\label{tensorproduct}
(a_1\otimes a_2)\cdot(a_1'\otimes a_2')=(a_1\cdot a_1')\otimes (a_2\cdot a_2'). 
\end{equation}
If the algebras $A_1,A_2$ are $\ZZ_2$-graded, $A_1\otimes A_2$ is also $\ZZ_2$-graded in an obvious way. On the other hand, for the supertensor product the multiplication is defined as follows:
\begin{equation}\label{supertensor}
(a_1\fotimes a_2)\cdot(a_1'\fotimes a_2')=(-1)^{\vert a_2\vert\cdot \vert a_1'\vert}(a_1\cdot a_1')\fotimes (a_2\cdot a_2'),
\end{equation}
where $(-1)^{\vert a\vert}$ is the fermionic parity of $a$. 

To derive (\ref{fermstacking}), we first note that
\begin{equation}
Z_\text{bose}(A_1,\beta_1) Z_\text{bose}(A_2,\beta_2)=Z_\text{bose}(A_1\otimes A_2,\beta_1,\beta_2),
\end{equation}
where we used the fact that the stacking of two bosonic systems with symmetry $\ZZ_2$ has a symmetry $\ZZ_2\times\ZZ_2$ and thus can be coupled to a pair of $\ZZ_2$ gauge fields $\beta_1,\beta_2$.
Next, it is easy to see that
\begin{equation}
Z(A_1\fotimes A_2,\beta_1,\beta_2)=(-1)^{\langle [\beta_1],[\beta_2]\rangle}Z(A_1\otimes A_2,\beta_1,\beta_2).
\end{equation}
These two identities together with (\ref{quad}) imply (\ref{fermstacking}).

As an illustration, consider $A=\Cl(1)$. Since apart from $1$ this algebra has a single odd basis element $\gamma$, $\beta$ completely determines the coloring of $\Gamma$. With the proper normalization of $Z_\text{bose}$, one gets
\begin{equation}
Z_\text{ferm}(s)=2^{-b_1(\Sigma)/2}\sum_{[\beta]} \sigma_s([\beta]).
\end{equation}
The r.h.s. is called the Arf invariant of the spin structure $s$ and is denoted $\Arf(s)$. One can show that it takes values $\pm 1$. If we stack two such systems together, we will get the partition function which is $1$ for all spin structures and all $\Sigma$, i.e. a trivial spin-TQFT.

It is easy to see that $\Cl(1)\fotimes \Cl(1)$ is the Clifford algebra with two generators, $\Cl(2)$. This algebra is non-trivial, but it is Morita-equivalent to the trivial algebra $\CC$. One can show that, just as in the bosonic case \cite{KTY}, spin-TQFT constructed from $A$ depends only on the Morita equivalence class of $A$. This explains why the spin-TQFT corresponding to $\Cl(2)$ is trivial.

We see that $A=\Cl(1)$ corresponds to a nontrivial SRE phase in the fermionic case (it is its own inverse). On the other hand, $\Cl(1)\otimes \Cl(1)$ is a commutative algebra isomorphic to a sum of two copies of $\Cl(1)$. Therefore the bosonic phase corresponding to $\Cl(1)$ is not invertible. This example illustrates that bosonization does not preserve the stacking operation.



\subsection{Including boundaries}


When $\Sigma$ has a non-empty boundary, $\Gamma$ is allowed to have univalent vertices which all lie on the boundary $\partial\Sigma$. Let $M$ be the number of boundary vertices. For every vertex $v$ we color each element of $\Gamma(v)$ with a basis vector of $A$, so that a vertex on the boundary has only a single label. As before, the weight of each coloring is a product of three factors: the product of $C_{ijk}$ over all trivalent vertices and $\eta^{ij}$ over all edges, the Koszul sign, and the spin-dependent sign. When summing over colorings, the labels of the  boundary vertices remain fixed. The result of the summation can be interpreted as a value of a map
\begin{equation}
Z_{\Gamma}(\Sigma): A^{\otimes M}\ra\CC,
\end{equation}
on a particular basis vector in $A^{\otimes M}$.

It is implicit here that the map depends on the spin structure on every connected component of $\partial\Sigma$. It can be read off from the function $\sigma_s(\gamma)$ evaluated on the boundary components. The spin structure is Neveu-Schwarz if $\sigma_s=1$ and Ramond if $\sigma_s=-1$. 



We can also consider open-closed spin-TQFT, i.e. spin-TQFT in the presence of topological boundary conditions (branes). Such boundary conditions are encoded in $\ZZ_2$-graded modules over $A$. A $\ZZ_2$-graded module over a $\ZZ_2$-graded algebra $A$ is a $\ZZ_2$-graded vector space $U=U_+\oplus U_-$ with the structure of an $A$-module such that $A_+\cdot U_\pm\subseteq U_\pm$ and $A_-\cdot U_\pm\subseteq U_\mp$. Equivalently, $U$ is an $A$-module equipped with an involution $P$ such that $T(\cF(a))=PT(a)P^{-1}$.


For each boundary component of $\Sigma$, choose a $\ZZ_2$-graded $A$-module $U$ and a homogeneous basis $f^U_\mu$ of $U$. Label each boundary edge with a basis vector of $U$. The weight of the coloring is a product of the $C$'s and $\eta$'s and a sign $\sigma_s(\beta)$, as well as a module tensor $T^\mu{}_{\nu i}$ for each boundary vertex. The sign is computed as before as a product of the spin-structure-dependent sign and the Koszul sign.

\section{Fermionic MPS}\label{MPSsection}

\subsection{Fermionic Matrix Product States and the annulus diagram}\label{MPSfromTQFT}

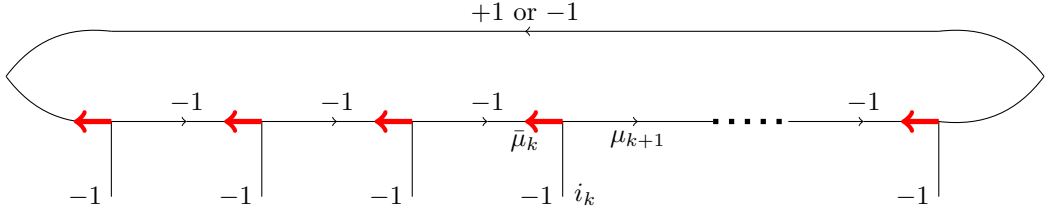
\begin{figure}
\centering

\begin{tikzpicture}[scale =2]

\draw (0,0) to[bend left] (-0.7,0.3);
\draw (-0.7,0.3) to[bend left] (0,0.6);
\begin{scope}
\draw[->] (0,0)--(0.5,0) node[above]{$-1$};
\draw (0.5,0)--(1,0);
\draw(1,0)--(1,-0.5) node[left]{$-1$};
\draw[red,line width=0.7mm][->] (1,0)--(0.75,0);
\end{scope}

 \begin{scope}[shift={(1,0)}]
\draw[->] (0,0)--(0.5,0) node[above]{$-1$};
\draw (0.5,0)--(1,0);
\draw(1,0)--(1,-0.5) node[left]{$-1$};
\draw[red,line width=0.7mm][->] (1,0)--(0.75,0);
\end{scope}

\begin{scope}[shift={(-1,0)}]
\draw(1,0)--(1,-0.5) node[left]{$-1$};
\draw[red,line width=0.7mm][->] (1,0)--(0.75,0);
\end{scope}

 \begin{scope}[shift={(2,0)}]
\draw[->] (0,0)--(0.5,0) node[above]{$-1$};
\draw (0.5,0)--(1,0);
\draw(1,0)--(1,-0.5) node[left]{$-1$};
\draw[red,line width=0.7mm][->] (1,0)--(0.75,0);
\draw (0.75,0) node[below]{$\bar{\mu}_k$};
\draw (1.15,-0.35) node[below]{$i_k$};
\draw[->] (1,0)--(1.5,0) node[below]{$\mu_{k+1}$};
\draw(1.5,0)--(2,0);
\draw[loosely dotted, line width = 0.7mm](2,0)--(2.5,0);
\end{scope}

 \begin{scope}[shift={(4.5,0)}]
\draw[->] (0,0)--(0.5,0) node[above]{$-1$};
\draw (0.5,0)--(1,0);
\draw(1,0)--(1,-0.5) node[left]{$-1$};
\draw[red,line width=0.7mm][->] (1,0)--(0.75,0);
\end{scope}

\draw (5.5,0) to [bend right] (6.2,0.3);
\draw (6.2,0.3) to [bend right] (5.5,0.6);

\draw[->](5.5,0.6)--(2.75,0.6) node[above] {$+1$ or $-1$};
\draw (2.75,0.6)--(0,0.6);

\end{tikzpicture}
\caption{Black arrows are edge orientations, and red arrows are special edges. All of the spin signs are $-1$ except possibly the one on the $N$-to-$1$ edge, which is $+1$ in the NS sector and $-1$ in the R sector.}
\label{fig:MPS}
\end{figure}

In this section, we will extract MPS wavefunctions from the spin-TQFT by considering the special case when $\Sigma$ is an annulus. Take one of the boundary circles to be a source cut boundary and the other to be a brane boundary corresponding to a $\ZZ_2$-graded $A$-module $U$ with action $T(a)\in\End(U)$. Choose a triangulation of $\Sigma$. It was shown in \cite{KTY} that one can deform the skeleton to look like Figure \ref{fig:MPS}.

Give the skeleton a marking and spin signs that models the spin structure on $\Sigma$. It is convenient to make the choices shown in Figure \ref{fig:MPS}. The sign on the $N$-to-$1$ edge is $+1$ if the spin structure induced on the boundary circles is NS and $-1$ if it is R. To get the sign \eqref{spinsign}, we insert a factor of $P$ for each $+1$.

Following the procedure detailed in Section \ref{spintqft} to evaluate the diagram in Figure \ref{fig:MPS}, one finds
\begin{equation}
Z(\Sigma_{T,NS})=\sum_{I=\{i_k,\mu_k,\bar\mu_k\}}\sigma_0(\beta_I)\times T^{\bar\mu_Ni_1\mu_1}T^{\bar\mu_1i_2\mu_2}\cdots T^{\bar\mu_{N-1}i_N\mu_N}\delta_{\mu_1\bar\mu_1}\delta_{\mu_2\bar\mu_2}\cdots P_{\mu_N\bar\mu_N}\bra{i_1i_2\cdots i_N}
\end{equation}in the NS sector and
\begin{equation}Z(\Sigma_{T,R})=\sum_{I=\{i_k,\mu_k,\bar\mu_k\}}\sigma_0(\beta_I)\times T^{\bar\mu_Ni_1\mu_1}T^{\bar\mu_1i_2\mu_2}\cdots T^{\bar\mu_{N-1}i_N\mu_N}\delta_{\mu_1\bar\mu_1}\delta_{\mu_2\bar\mu_2}\cdots\delta_{\mu_N\bar\mu_N}\bra{i_1i_2\cdots i_N}
\end{equation}in the R sector, where the Koszul sign is given as a Grassmann integral
\begin{align}\sigma_0(\beta_I)=\int d\theta_1^{|\mu_1|}d\bar\theta_1^{|\bar\mu_1|}d\theta_2^{|\mu_2|}d\bar\theta_2^{|\bar\mu_2|}\cdots d\theta_N^{|\mu_N|}d\bar\theta_N^{|\bar\mu_N|}d\theta_{i_1}^{|i_1|}d\theta_{i_2}^{|i_2|}\cdots d\theta_{i_N}^{|i_N|}\nonumber\\\times\quad\bar\theta_N^{|\bar\mu_N|}\theta_{i_1}^{|i_1|}\theta_1^{|\mu_1|}\bar\theta_1^{|\bar\mu_1|}\theta_{i_2}^{|i_2|}\theta_2^{|\mu_2|}\cdots\bar\theta_{N-1}^{|\bar\mu_{N-1}|}\theta_{i_N}^{|i_N|}\theta_N^{|\mu_N|}\end{align}Evaluating the integral amounts to reordering the variables in the integrand to match the ordering in the measure while recording the sign
\begin{equation}\theta_1^{s_1}\theta_2^{s_2}=(-1)^{s_1s_2}\theta_2^{s_2}\theta_1^{s_1}
\end{equation}Moving $\bar\theta_N^{|\bar\mu_N|}$ across the integrand gives a sign $(-1)^{|\bar\mu_N|}$. Then moving each $\theta_{i_k}^{|i_k|}$ to the right gives a sign $+1$. Therefore the total sign is
\begin{equation}
\sigma_0(\beta_I)=(-1)^{|\bar\mu_N|}
\end{equation}Noting that $\delta_{\mu_n\bar\mu_N}(-1)^{|\bar\mu_N|}=P_{\mu_n\bar\mu_N}$, we find that the MPS wavefunctions take the forms
\begin{equation}\label{NSMPS1}\bra{\psi_{T,NS}}=Z(\Sigma_{T,NS})=\sum_{i_1,i_2,\cdots,i_N}\Tr[T(e_{i_1})T(e_{i_2})\cdots T(e_{i_N})]\bra{i_1i_2\cdots i_N}
\end{equation}and
\begin{equation}\label{RMPS1}\bra{\psi_{T,R}}=Z(\Sigma_{T,R})=\sum_{i_1,i_2,\cdots,i_N}\Tr[PT(e_{i_1})T(e_{i_2})\cdots T(e_{i_N})]\bra{i_1i_2\cdots i_N}
\end{equation}

More general states, called \emph{generalized MPS}, on the closed chain are obtained from the spin-TQFT by inserting a local observable on the brane boundary of the annulus. Such observables are parametrized by linear maps $X:U\rightarrow U$ and can be either even or odd; that is, $PX=XP$ or $PX=-XP$, respectively.

The NS sector MPS resulting from the insertion of $X$ has conjugate wavefunction
\begin{equation}\label{NSMPS}\bra{\psi_{T,NS}^X}=
\sum_{i_1\cdots i_N}\tr[X^\dagger T(e_{i_1})\cdots T(e_{i_N})]\bra{i_1
\cdots i_N}
\end{equation}In the R sector,
\begin{equation}
\label{RMPS}\bra{\psi_{T,R}^X}=
\sum_{i_1\cdots i_N}\tr[PX^\dagger T(e_{i_1})\cdots T(e_{i_N})]\bra{i_1
\cdots i_N}
\end{equation}
Note that the generalized MPS corresponding to the trivial observable $X=\mathds{1}$ are the states $\bra{\psi_{T}}$ \eqref{NSMPS1}\eqref{RMPS1}.

The state $\ket{\psi_{T,NS/R}^X}$ has the same fermionic parity as the observable $X$ since
\begin{align}\label{sameparity}
\cF^{\otimes N}\bra{\psi_{T,NS(R)}^X} = \sum \Tr[(P)X^{\dagger} T(\cF\cdot e_{i_1}) \cdots T(\cF\cdot e_{i_n})] \langle i_1 \cdots i_n | \nonumber \\ 
= \sum \Tr[(P)PX^{\dagger}PT(e_{i_1})\cdots T(e_{i_n}) ] \langle i_1 \cdots i_n | \nonumber \\
= (-1)^{|X|}\bra{\psi_{T,NS(R)}^X}
\end{align}

\subsection{Parent Hamiltonians}\label{hamsection}

\begin{figure}
\centering
\begin{subfigure}[b]{0.5\textwidth}
\centering
\begin{tikzpicture}[scale=1.2]

\draw (0,0) ellipse (1.2 and 0.2);
\draw (0.3,-0.2) -- (0.3,-1);
\draw (-0.3,-0.2) -- (-0.3,1);

\draw (1.6,0) -- (4,0);
\draw (3.1,0) -- (3.1,-1);
\draw (2.5,0) -- (2.5,1);

\end{tikzpicture}
\caption{Skeleton of an annulus with cut boundaries}
\end{subfigure}%
\begin{subfigure}[b]{0.5\textwidth}
\centering
\begin{tikzpicture}[scale=1.1]

\draw (0,1) ellipse (1.2 and 0.2);
\draw (0,0) ellipse (1.2 and 0.2);
\draw (0.3,-0.2) -- (0.3,-1);
\draw (-0.3,-0.2) -- (-0.3,0.8);

\draw (1.6,1) -- (4,1);
\draw (1.6,0) -- (4,0);
\draw (3.1,0) -- (3.1,-1);
\draw (2.5,0) -- (2.5,1);

\end{tikzpicture}
\caption{An annulus with one brane and one cut boundary}
\end{subfigure}
\caption{}
\label{fig:Ham}
\end{figure}

Hamiltonians appear in TQFT as cylinders. There is one for each of the NS and R sectors. To be precise, the Hamiltonian is the linear map
\begin{equation}
H_{NS(R)}=\mathds{1}-Z(C_{NS(R)})
\end{equation}where $C_{NS(R)}$ denotes the cylinder with NS (R) spin structure. The composition of two cylinder cobordisms is again a cylinder, so $Z(C)$ is a projector, and therefore so is $H$. Ground states are those with eigenvalue $1$ under $Z(C)$. It is convenient to specialize to the case of a single site, $N=1$. Since these Hamiltonians arise from a topologically-invariant theory, properties of the $N=1$ system must hold more generally. Consider the skeleton of the cylinders depicted in Figure \ref{fig:Ham}.




\begin{figure}
\centering

$Z(C)$\hspace{5mm}$=$\hspace{5mm}$\frac12\sigma_1$
\begin{tikzpicture}[baseline={([yshift=-.5ex]current bounding box.center)}]
\centering

\draw(2,0)--(2,-1);
\draw(1,-1)--(4,-1);
\draw(3,-1)--(3,-2);
\end{tikzpicture}\hspace{5mm}$+$\hspace{5mm}$\frac12\sigma_2$ \begin{tikzpicture}[baseline={([yshift=-.5ex]current bounding box.center)}]
\centering

\draw(2,0)--(2,-1);
\draw[magenta,line width=0.5mm](1,-1)--(4,-1);
\draw(3,-1)--(3,-2);
\end{tikzpicture}

\hspace{5mm}$+$\hspace{5mm}$\frac12\sigma_3$ \begin{tikzpicture}[baseline={([yshift=-.5ex]current bounding box.center)}]
\centering

\draw[magenta,line width=0.5mm](2,0)--(2,-1);
\draw(1,-1)--(2,-1);
\draw[magenta,line width=0.5mm](2,-1)--(3,-1);
\draw(3,-1)--(4,-1);
\draw[magenta,line width=0.5mm](3,-1)--(3,-2);
\end{tikzpicture}\hspace{5mm}$+$\hspace{5mm}$\frac12\sigma_4$ \begin{tikzpicture}[baseline={([yshift=-.5ex]current bounding box.center)}]
\centering

\draw[magenta,line width=0.5mm](2,0)--(2,-1);
\draw[magenta,line width=0.5mm](1,-1)--(2,-1);
\draw(2,-1)--(3,-1);
\draw[magenta,line width=0.5mm](3,-1)--(4,-1);
\draw[magenta,line width=0.5mm](3,-1)--(3,-2);
\end{tikzpicture}

\caption{The cylinder partition sum $Z(C)$ factors as a signed sum of four colored diagrams: $\sigma(\beta_1)C_1+\sigma(\beta_2)C_2+\sigma(\beta_3)C_3+\sigma(\beta_4)C_4=C_1+\eta C_2+C_3-\eta C_4$. Magenta lines indicate odd edges.}
\label{fig:hamcolor}
\end{figure}

\begin{figure}
\centering
$\bra{\psi_\text{even}}$\hspace{5mm} $=$ \hspace{5mm}$\sigma_1$ 
\begin{tikzpicture}[baseline={([yshift=-.5ex]current bounding box.center)}]
\centering

\path (0,0) (1,0) node[circle,inner sep=1pt, draw](X){$X$};

\draw(0,0)--(X);
\draw(X)--(3,0);
\draw(2,0)--(2,-1);
\end{tikzpicture}\hspace{5mm}$+$\hspace{5mm}$\sigma_2$
\begin{tikzpicture}[baseline={([yshift=-.5ex]current bounding box.center)}]
\centering

\path (0,0) (1,0) node[circle,inner sep=1pt, draw](X){$X$};

\draw[magenta,line width=0.5mm](0,0)--(X);
\draw[magenta,line width=0.5mm](X)--(3,0);
\draw(2,0)--(2,-1);
\end{tikzpicture}

\caption{$\bra{\psi_\text{even}}=\sigma(\beta_1)\bra{\psi_1}+\sigma(\beta_2)\bra{\psi_2}=\bra{\psi_1}+\eta\bra{\psi_2}$}
\label{fig:evenmpscolor}
\end{figure}

\begin{figure}
\centering
$\langle \psi_{\rm{odd}}|$ \hspace{5mm} $=$ \hspace{5mm}$\sigma_3$ 
\begin{tikzpicture}[baseline={([yshift=-.5ex]current bounding box.center)}]
\centering

\path (0,0) (1,0) node[circle,inner sep=1pt, draw](X){$X$};

\draw(0,0)--(X);
\draw[magenta,line width=0.5mm](X)--(2,0);
\draw(2,0)--(3,0);
\draw[magenta,line width=0.5mm](2,0)--(2,-1);
\end{tikzpicture}\hspace{5mm}$+$\hspace{5mm}$\sigma_4$
\begin{tikzpicture}[baseline={([yshift=-.5ex]current bounding box.center)}]
\centering

\path (0,0) (1,0) node[circle,inner sep=1pt, draw](X){$X$};

\draw[magenta,line width=0.5mm](0,0)--(X);
\draw(X)--(2,0);
\draw[magenta,line width=0.5mm](2,0)--(3,0);
\draw[magenta,line width=0.5mm](2,0)--(2,-1);
\end{tikzpicture}

\caption{$\bra{\psi_\text{odd}}=\sigma(\beta_3)\bra{\psi_3}+\sigma(\beta_4)\bra{\psi_4}=\bra{\psi_3}+\eta\bra{\psi_4}$}
\label{fig:oddmpscolor}
\end{figure}

By exploiting \eqref{fermstatesum}, we will not need the full machinery of lattice spin structures to understand the Hamiltonians and their ground states. The path integrals for the cylinders can be expressed as a sum over the four relative $1$-cycles $\beta_1,\ldots,\beta_4$ depicted in Figure \ref{fig:hamcolor}. The first colored diagram corresponds to the trivial cycle $\beta_1$ and has no odd labels, so its sign is trivial, $\sigma_s(\beta_1)=1$. The second one corresponds to the equator of the cylinder and comes with the sign $\sigma(\beta_2):=\eta$, which is $+1$ in the NS sector and $-1$ in the R sector. The relative cycles $\beta_3$ and $\beta_4$ sum to $\beta_2$ and have intersection number $1$, where the intersection pairing is defined by gluing another annulus onto the annulus, to get a torus $C^*=T^2$, as explained in Section \ref{spinstructsection}. Therefore \eqref{quad} says there is a relative sign
\begin{equation}
\sigma_s(\beta_3)\sigma_s(\beta_4)=\sigma_s(\beta_3+\beta_4)(-1)^{\langle\beta_3,\beta_4\rangle}=\sigma_s(\beta_2)(-1)=-\eta
\end{equation}One can choose a spin structure on the closed space $C^*=T^2$ such that $\sigma_s(\beta_3)=1$; this amounts to fixing trivializations of the spin structures induced on each component of $\partial C$ at the univalent vertices.

Similarly, an even MPS can be expressed as the sum in Figure \ref{fig:evenmpscolor}, where $\sigma_1=1$ and $\sigma_2=\eta$, and an odd MPS as the sum in Figure \ref{fig:oddmpscolor} with $\sigma_1=1$ and $\sigma_2=\eta$.

Now we are ready to argue that the parent Hamiltonian has a generalized MPS $\bra{\psi_T^X}$ a ground state if $X$ \emph{supercommutes} with $T(a)$; that is, if an even observable satisfies
\begin{equation}
\label{evenobs}XT(a)=T(a)X\quad\forall a\in A,
\end{equation}and an odd observable satisfies
\begin{equation}
\label{oddobs}XT(a)=(-1)^{|a|}T(a)X\quad\forall a\in A.
\end{equation}Linear maps satisfying these conditions are called even and odd $\ZZ_2$-graded module endomorphisms.

\begin{figure}
$C_1(\psi_1)=$
\begin{tikzpicture}[baseline={([yshift=-.5ex]current bounding box.center)}]
\centering

\path (0,0) (1,0) node[circle,inner sep=1pt, draw](X){$X$};

\draw(0,0)--(X);
\draw(X)--(3,0);
\draw(2,0)--(2,-1);
\draw(1,-1)--(4,-1);
\draw(3,-1)--(3,-2);

\end{tikzpicture} $=$\hspace{1cm}\begin{tikzpicture}[baseline={([yshift=-.5ex]current bounding box.center)}]

\centering

\path (0,0) (1,0) node[circle,inner sep=1pt, draw](X){$X$};

\draw (0,0)--(X);
\draw (X)--(2,-0.5);
\draw (3,-0.5)--(4,0)--(5,0);
\draw (2,-0.5)--(3,-0.5);
\draw(2,-0.5)--(1,-1)--(0,-1);
\draw (3,-0.5)--(4,-1)--(5,-1);
\draw (4.5,-1)--(4.5,-2);

\end{tikzpicture} \hspace{1cm} $=$ \begin{tikzpicture}[baseline={([yshift=-.5ex]current bounding box.center)}]

\centering

\path (2,0.5) node[circle,inner sep=1pt, draw](X){$X$};

\draw (0,0) -- (1,0);
\draw(1,0)--(1.5,0.5)--(X);
\draw(X)--(2.5,0.5)--(3,0);
\draw(3,0)--(4,0);
\draw(1,0)--(2,-0.5)--(3,0);
\draw(2,-0.5)--(2,-1.5);
\end{tikzpicture} \hspace{0.5cm}$= $ \hspace{1cm} \begin{tikzpicture}[baseline={([yshift=-.5ex]current bounding box.center)}]

\centering

\path (1,0) node[circle,inner sep=1pt, draw](X){$X$};
\draw (0,0) -- (X);
\draw (X)--(2,0);
\draw(2,0)--(3,0);
\draw(3,0)--(4,0);
\draw(2,0)--(2.5,-1)--(3,0);
\draw (2.5,-1)--(2.5,-2);

\end{tikzpicture} $=$\hspace{1cm}\begin{tikzpicture}[baseline={([yshift=-.5ex]current bounding box.center)}]

\centering

\path (1,0) node[circle,inner sep=1pt, draw](X){$X$};
\draw(0,0) -- (X)--(4,0);
\draw (3,0)--(3,-1);
\end{tikzpicture}

\vspace{0.5cm} $= \psi_1$
\caption{Diagrammatic proof of $C_1 \langle \psi_1| = \langle \psi_1|$. The topmost line represents the physical boundary, with module indices living on it. The others are depicted in Appendix \ref{sec:diagrams}.}
\label{fig:p1}
\end{figure}
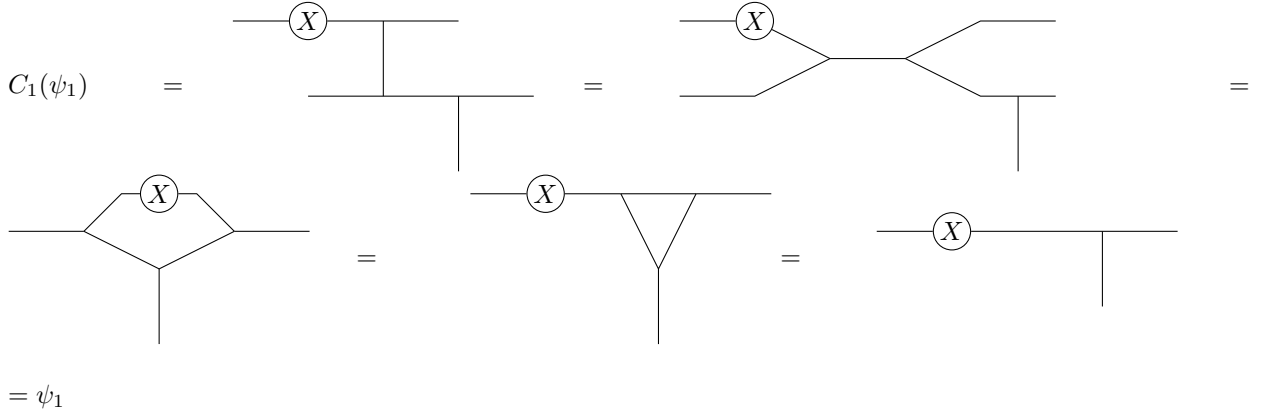

The maps $C_3$ and $C_4$ correspond to diagrams with odd legs, and so annihilate even states $\bra{\psi_\text{even}}$. Therefore
\begin{equation}
Z(C)\bra{\psi_{\text{even}}}=\tfrac{1}{2}(C_1+\eta C_2)(\bra{\psi_1}+\eta\bra{\psi_2})
\end{equation}By the sequence of diagram moves depicted in Figures \ref{fig:p1}, \ref{fig:p2}, \ref{fig:p3}, and \ref{fig:p4}, one can show that
\begin{equation}
C_1\bra{\psi_1}=\bra{\psi_1},\quad C_2\bra{\psi_1}=\eta_X\bra{\psi_2},\quad C_1\bra{\psi_2}=\bra{\psi_2}, \quad C_2\bra{\psi_2}=\eta_X\bra{\psi_1}
\end{equation}where $\eta_X$ denotes the sign due to commuting $X$ with odd $T(a)$. According to the rule \eqref{evenobs}, $\eta_X=1$, so
\begin{equation}
Z(C)\bra{\psi_{\text{even}}}=\tfrac{1}{2}(1+\eta_X)\bra{\psi_{\text{even}}}=\bra{\psi_{\text{even}}}
\end{equation}Similarly, the cylinder acts on odd states as 
\begin{equation}
Z(C)\bra{\psi_\text{odd}}=\tfrac{1}{2}(C_3-\eta C_4)(\bra{\psi_3}+\eta\bra{\psi_4})
\end{equation}Commuting $X$ with the vertex gives $\bra{\psi_4}=\eta_X\bra{\psi_3}$, which means $\bra{\psi_{\text{odd}}}=(1+\eta\eta_X)\bra{\psi_3}$. According to the rule \eqref{oddobs}, $\eta_X=-1$, so the only odd ground state in the NS sector is $\bra{\psi}=0$. This agrees with \cite{MS}. 

In the Ramond sector, one can have nonzero odd ground states. The sequence of moves of Figures \ref{fig:p5} and \ref{fig:p6} shows
\begin{equation}
C_3\bra{\psi_3}=\bra{\psi_3},\quad C_4\bra{\psi_3}=\bra{\psi_3}
\end{equation}so
\begin{equation}
Z(C)\bra{\psi_{\text{odd}}}=\tfrac{1}{2}(1-\eta)\bra{\psi_{\text{odd}}}=\bra{\psi_{\text{odd}}} \text{ (in the R sector)}
\end{equation} 
Therefore $\bra{\psi_T^X}$ is indeed a ground state of $H_{NS(R)}$ provided $X$ is a $\ZZ_2$-graded module endomorphism.

Next we argue that every ground state of $H$ of the form \eqref{NSMPS} or \eqref{RMPS} for arbitrary $X$ can be written as a generalized MPS where $X$ supercommutes with $T$. A result of Ref. \cite{MS} (c.f. eq 3.18) implies that
\begin{equation}\label{ms1}
Z(C_{NS})\ket{ij}=(-1)^{|i||j|+|i|}Z(C_{NS})\ket{ji}
\end{equation}and\begin{equation}\label{ms2}Z(C_R)\ket{ij}=(-1)^{|i||j|}Z(C_R)\ket{ji}.\end{equation}
In Appendix \ref{MSproof}, we rederive this result in the Novak-Runkel formalism. Then, since $|X|=|i|+|j|$,
\begin{equation}
Z(C_{NS})\Tr[XT(e_i)T(e_j)]\ket{ij}=(-1)^{|i||X|}Z(C_{NS})\Tr[T(e_i)XT(e_j)]\ket{ji}
\end{equation}and
\begin{equation}
Z(C_R)\Tr[PXT(e_i)T(e_j)]\ket{ij}=(-1)^{|i||X|}Z(C_R)\Tr[PT(e_i)XT(e_j)]\ket{ji}.
\end{equation}
For ground states, i.e. eigenstates of $Z(C)$ with eigenvalue $1$, this means that $X$ supercommutes with $T$.


It turns out that all ground states of $H$ can be written as generalized MPS. As discussed in \cite{KTY}, in a unitary theory $T$ is an isometry with respect to some inner product on $A$ and the standard inner product
\begin{equation}\langle M|N\rangle =\Tr[M^\dagger N]\quad M,N\in\End(V)
\end{equation}on $\End(V)$. For an orthogonal basis $\{e_i\}$ of $A$, $\Tr[T(e_i)^\dagger T(e_j)]=\delta_{ij}$. Consider the case $N=1$. An arbitrary state
\begin{equation}\bra{\psi}=\sum_i a_i\bra{i}
\end{equation}can be written in generalized MPS form \eqref{NSMPS}\eqref{RMPS} if one takes
\begin{equation}X_{NS}=\sum_ja_jT(e_j)^\dagger\quad\text{ or }\quad X_R=\sum_ja_jPT(e_j)^\dagger.
\end{equation}Thus generalized MPS with supercommuting $X$ are the only ground states. Neither the number of generalized MPS nor the number of ground states depends on $N$; thus, the argument extends to all $N$.

A consequence of supercommutativity and \eqref{sameparity} is that there are no odd ground states in the NS sector. Suppose that $X$ is an odd observable. For $a\in A_-$, the matrix $X^\dagger$ anticommutes with $T(a)$, so the coefficient $\Tr[X^\dagger T(a)]$ vanishes. For $a\in A_+$, the matrix $X^\dagger T(a)$ maps $U_\pm$ to $U_\mp$ and so also vanishes in the trace. Therefore the state \eqref{NSMPS} is zero for odd $X$, which is to say that the NS sector does not support odd states. The argument fails for the state \eqref{RMPS}; generically, the R sector supports both even and odd states. The lack of odd states in the NS sector can also be seen directly from \eqref{ms1}, which implies $|C\ket{ij}|=|i|+|j|=0$.





\subsection{Stacking fermionic MPS}

Bosonization establishes a 1-1 correspondence between 1d bosonic systems with $\ZZ_2$ symmetry and 1d fermionic systems. In the gapped case, the corresponding topological phases are described by the same algebraic data, namely by a $\ZZ_2$-graded algebra $A$. But bosonization does not preserve a crucial physical structure: stacking systems together. From the mathematical viewpoint, either bosonic or fermionic topological phases of matter form a commutative monoid (a set with a commutative associative binary operation and a neutral element, but not necessarily with an inverse for every element), but bosonization does not preserve the monoid structure (i.e. it does not preserve the product). A well-known example is given by the fermionic SRE phases: the non-trivial fermionic SRE phase (the Majorana chain) is mapped to the bosonic phase with a spontaneously broken $\ZZ_2$. The former one is invertible, while the latter one is not. Both phases correspond to the  algebra $\Cl(1)$.

In the bosonic case, it was shown in \cite{KTY} that, given two algebras $A_1$ and $A_2$ with bosonic Hamiltonians $H_1$ and $H_2$, the tensor product system $A_1\otimes A_2$ has a Hamiltonian $H_1\otimes\mathds{1}_2+\mathds{1}_1\otimes H_2$. That is, stacking bosonic systems together corresponds to the tensor product of algebras.

On the other hand, in section 2.4 we have shown that for fermionic systems stacking corresponds to the supertensor product (\ref{supertensor}). We can now see that the supertensor product rule is consistent with the way fermionic generalized MPS are defined (while the usual tensor product is not).  


Suppose $H_1$ is the Hamiltonian for the MPS system built from a $\ZZ_2$-graded algebra $A_1$ that acts on a $\ZZ_2$-graded module $U_1$ by $T_1$. Its ground states are parametrized by $\ZZ_2$-graded module endomorphisms $X_1$ of $U_1$. Consider stacking $H_1$ with a second system $H_2$ defined by $T_2:A_2\rightarrow\End(U_2)$ with ground states parametrized by $X_2$. The stacked system is the MPS system with physical space $A_1\otimes A_2$ and Hamiltonian $H=H_1\otimes\mathds{1}_2+\mathds{1}_1\otimes H_2$. It has bond space $U_1\otimes U_2$ and MPS tensor $T=T_1\otimes T_2$.

The ground states are generalized MPS, and so correspond to $\ZZ_2$-graded endomorphisms of the module $U_1\otimes U_2$. Since the MPS tensor is $T=T_1\otimes T_2$, the state $\bra{\psi_T^X}$ is trivial unless $X$ is of the form $X_1\otimes X_2$. We also know that $X$ supercommutes with $T$:\begin{equation}\label{stacksupcom}(X_1\otimes X_2)(T_1\otimes T_2)=(-1)^{(|X_1|+|X_2|)(|T_1|+|T_2|)}(T_1\otimes T_2)(X_1\otimes X_2)
\end{equation}

There are two ways one might define the composition of tensor products of operators\footnote{These correspond to the two symmetric monoidal structures on the category of $\ZZ_2$-graded vector spaces.}:
\begin{equation}\label{bosXT}(X_1\otimes X_2)(T_1\otimes T_2)=X_1T_1\otimes X_2T_2
\end{equation}and
\begin{equation}\label{fermXT}(X_1\fotimes X_2)(T_1\fotimes T_2)=(-1)^{|X_2||T_1|}X_1T_1\fotimes X_2Y_2
\end{equation}
Since $X_1$ supercommutes with $T_1$ and $X_2$ with $T_2$, only the second notion \eqref{fermXT} of composition is consistent with \eqref{stacksupcom}. The composition rule is an algebra structure on $\End(U_1)\otimes\End(U_2)$ and pulls back by $T$ to an algebra structure on $A_1\otimes A_2$ given by the rule (\ref{supertensor}).

An important assumption in this argument is that isomorphic TQFTs correspond to equivalent gapped phases. Assuming this is true, we can easily see that the group of fermionic SRE phases is isomorphic to $\ZZ_2.$ Indeed, one can easily see that a phase which is invertible must correspond to an indecomposable algebra (i.e. the algebra which cannot be decomposed as a sum of algebras). Since all our algebras are semisimple, this means that invertible phases must correspond to simple algebras. It is well-known that there are exactly two Morita-equivalence classes of $\ZZ_2$-graded algebras: the trivial one and the class of $\Cl(1)$. The square of the nontrivial class is the trivial class. Hence the group of invertible fermionic phases is isomorphic to $\ZZ_2$. In the next section we will show explicitly that $\Cl(1)$ corresponds to the nontrivial Majorana chain.

\section{Hamiltonians for fermionic SRE phases}

\subsection{The trivial SRE phase}\label{triv}

An example of a system in the trivial phase is the trivial Majorana chain \cite{fidkit}. On a circle, this system has only bosonic states: one in the NS sector and one in the R sector. We will now demonstrate that this is the same phase as the MPS system built out of the Clifford algebra $\Cl(2)=\End(\CC^{1|1})$.

The algebra $A=C\ell(2)$ is expressed in terms of its odd generators as $\CC[x,y]/(x^2-1,y^2-1,xy+yx)$. Let $A$ act on $U=\CC^{1|1}$ by
\begin{equation}T:x\mapsto[\sigma_x]_\pm\quad\text{ , }\quad y\mapsto[\sigma_y]_\pm
\end{equation}where $[\cdot]_\pm$ denotes a matrix in the homogeneous basis of $U$. This action is graded and faithful. The fermion parity operator $P$ acts by $\sigma_z$.

The even ground states of this system are parametrized by matrices that commute with $\sigma_x$, $\sigma_y$, and $\sigma_z$. Thus $X$ is proportional to the identity $\mathds{1}$. The corresponding NS sector state has the wavefunction $\Tr[T(e_{i_1})\cdots T(e_{i_N})]$. There is also an even state in the R sector given by $\Tr[PT(e_{i_1})\cdots T(e_{i_N})]$.

The odd ground states are parametrized by matrices that commute with $T(a)$ -- in particular, $T(xy)=\sigma_z$ -- and anticommute with $P=\sigma_z$. This is impossible, so there are no odd states in either sector.

In summary, the ground states of the $A=\Cl(2)$ MPS system are a bosonic one in the NS sector and a bosonic one in the R sector, just like the ground states of the trivial Majorana chain.

One can show that the MPS parent Hamiltonian (c.f. \cite{KTY,MPSone}) is a nearest-neighbor Hamiltonian with the two-body interaction $H_T=-\sum_{\alpha=1}^4\ket{v_\alpha}\bra{v_\alpha}$ where
\begin{align}v_1&=1\otimes 1-x\otimes x-y\otimes y-xy\otimes xy\nonumber\\v_2&=1\otimes x+x\otimes 1+y\otimes xy-xy\otimes y\nonumber\\v_3&=1\otimes y+y\otimes 1+xy\otimes x-x\otimes xy\nonumber\\v_4&=1\otimes xy+xy\otimes 1+x\otimes y-y\otimes x
\end{align}It is not obvious that $H_T$ is equivalent to the Hamiltonian of the trivial Majorana chain
\begin{equation}H=\sum_j (a_j^\dagger a_j-1)
\end{equation}but it should be possible to construct an LU transformation between the two Hamiltonians (after some  blocking), as the systems have the same spaces of ground states and so lie in the same phase.





\subsection{The nontrivial SRE phase}\label{nontriv}

An example of a fermionic system in a nontrivial SRE phase is the Majorana chain with a two-body Hamiltonian \cite{fidkit}
\begin{equation}\label{nontrivmaj}H_j=\frac{1}{2}\left(-a_j^\dagger a_{j+1}-a_{j+1}^\dagger a_j+a_j^\dagger a_{j+1}^\dagger+a_{j+1} a_j\right)
\end{equation} 
This system has one bosonic and one fermionic ground state on the interval arising from one Majorana zero mode at each end. In the continuum limit this system becomes a free Majorana fermion with a negative mass. In the NS sector there is a unique ground state which is bosonic, while in the R sector there is a unique ground state which is fermionic (this is most easily seen from the continuum field theory). 


In order to get this phase from a spin TQFT, we let $A=C\ell(1)$. To see the full space of ground states, we need a faithful graded module over $A$. Let $U=U_+\oplus U_-$, where each $U_\pm$ is spanned by a single vector $u_\pm$. Let $A$ act on $U$ by
\begin{equation}T:\Gamma\mapsto [\sigma_x]_\pm=u_+\otimes u_-^*+u_-\otimes u_+^*.
\end{equation} In other words, $U$ is $A$ regarded as a module over itself.

The even ground states of this system are parametrized by matrices that commute with $P=[\sigma_z]_\pm$ and $T(\Gamma)=[\sigma_x]_\pm$. Such matrices are proportional to $\mathds{1}$. The corresponding NS sector state has wavefunction $\Tr[T(e_{i_1})\cdots T(e_{i_N})]$. There is no even state in the R sector as the trace $\Tr[PT(e_1)\cdots T(e_{i_N})]$ vanishes.

The odd ground states are parametrized by matrices that anticommute with $P$ and $T(\Gamma)$. Such matrices $X$ are all proportional to $[\sigma_y]_\pm$. By the general argument of Section \ref{hamsection}, we know that the NS sector has no odd states. The wavefunction $\Tr[PX^\dagger T(e_{i_1})\cdots T(e_{i_N})]$ defines an odd state in the R sector.

In summary, the ground states of the $A=\Cl(1)$ MPS system are a bosonic one in the NS sector and a fermionic one in the R sector, just like the ground states of the nontrivial Majorana chain.

We can also observe the equivalence of the two systems from the standpoint of Hamiltonians. We build the MPS parent Hamiltonian for the $A=C\ell(1)$ system by following Ref. \cite{KTY,MPSone}. The adjoint $\cP=T^\dagger$ is given by
\begin{equation}\cP:2u_\pm\otimes u_\pm^*\mapsto 1\otimes 1+\Gamma\otimes\Gamma\quad\text{ , }\quad 2u_\pm\otimes u_\mp^*\mapsto 1\otimes\Gamma+\Gamma\otimes 1
\end{equation}With respect to the inner products on $A$ and $U$ for which $1$ and $\Gamma$ and $u_+$ and $u_-$ are unit vectors, the graded module structure $T$ is an isometry, so the left inverse $\cP^+$ is simply $T$. Putting these pieces together, we find
\begin{equation}\label{nontrivham}H_T=\ket{11}\bra{\Gamma\Gamma}-\ket{1\Gamma}\bra{\Gamma 1}-\ket{\Gamma 1}\bra{1\Gamma}+\ket{\Gamma\Gamma}\bra{11}
\end{equation}where $\ket{ab}\bra{cd}$ denotes the element $a\otimes b\otimes c^*\otimes d^*\in\End(A\otimes A)$. In terms of the annihilation operators $a_j=\sqrt{2}\ket{1}\bra{\Gamma}_j$ and their adjoints, the hopping (top row) and pairing (bottom) terms look like
\begin{eqnarray}a_j^\dagger\otimes a_{j+1}=2\ket{\Gamma 1}\bra{1\Gamma}&\hspace{1cm}&a_{j+1}^\dagger\otimes a_j=2\ket{1\Gamma}\bra{\Gamma 1}\nonumber\\a_j^\dagger\otimes a_{j+1}^\dagger=2\ket{\Gamma\Gamma}\bra{11}&\hspace{1cm}&a_{j+1}\otimes a_j=2\ket{11}\bra{\Gamma\Gamma}
\end{eqnarray}so the Hamiltonians \eqref{nontrivmaj} and \eqref{nontrivham} agree. The variables $a_j$ satisfy fermionic anti-commutation relations. For example,
\begin{equation}\{a_j,a_{j+1}\}=(a\otimes\mathds{1})(\mathds{1}\otimes a)+(\mathds{1}\otimes a)(a\otimes\mathds{1})=a\otimes a+(-1)^{|a||a|}a\otimes a=0
\end{equation}if we are careful to use the fermionic tensor product \eqref{supertensor}. The other relations can be checked similarly.

\section{Equivariant spin-TQFT and equivariant  fermionic MPS}




\subsection{$(\cG,p)$-equivariant algebras and modules}






Let $(\cG,p)$ be a finite supergroup, i.e. a finite group $\cG$ with a distinguished involution $p\in \cG$ called \emph{fermion parity}. We assume the involution $p$ is central in $\cG$, which means that there are no supersymmetries. Every supergroup $(G,p)$ arises as a central extension of a group $G_b\simeq\cG/\ZZ_2$ of bosonic symmetries by $\ZZ_2=\{1,p\}$; that is, there is an exact sequence
\begin{equation}\label{ses}1\rightarrow\ZZ_2\xrightarrow{i}\cG\xrightarrow{b} G_b\rightarrow 1.
\end{equation}
A trivialization of $(\cG,p)$ is a function $t:\cG\rightarrow\ZZ_2$ such that $t\circ i$ is the identity on $\ZZ_2$. Given a trivialization, one can encode the multiplication rule for $\cG$ in terms of the product on $G_b$ and a $\ZZ_2$-valued group $2$-cocycle $\rho$ of $G_b$. Consider the following product on the set $G_b\times\ZZ_2$ (denoted $G_b\times_\rho\ZZ_2$). For $\bar g,\bar h\in G_b, f,f^\prime\in\ZZ_2$,
\begin{equation}(\bar g,f)\cdot (\bar h,f^\prime)=(\bar g\bar h,\rho(\bar g,\bar h)+f+f^\prime)
\end{equation}
Denote $\bar g:=b(g)$. The map $b\times_\rho t:g\mapsto (\bar g,t(g))$ defines a group isomorphism $\cG\xrightarrow{\sim} G_b\times_\rho\ZZ_2$; that is,
\begin{equation}\label{extcocycle}g\cdot h=(\bar g,t(g))\cdot (\bar h,t(h))=(\bar g\bar h,\rho(\bar g,\bar h)+t(g)+t(h))=(\bar{gh},t(gh))=gh,
\end{equation}
if and only if
\begin{equation}\label{ctriv}\rho(\bar g,\bar h)=t(gh)+t(g)+t(h).
\end{equation}
Suppose $t^\prime$ is another trivialization. Since $t=t^\prime$ on the image of $i$ and the sequence \eqref{ses} is exact, the map $t-t^\prime$ defines a $1$-cochain of $G_b$. Thus, upon replacing $t$ with $t^\prime$, $\rho$ is modified by the coboundary $\delta(t-t^\prime)$, so only the cohomology class $[\rho]$ of $c$ is an invariant of the extension. If $[\rho]$ is trivial, $\cG$ is isomorphic to the direct product group $G_b\times\ZZ_2$ and we say the extension \emph{splits}; in general, this is not the case. Some discussions of fermionic phases in the physics literature assume that $(\cG,p)$ is split, but we will consider both cases simultaneously. Note that \cite{fidkit} considered both cases as well.


An action $R$ of $(\cG,p)$ on a vector space $V$ endows it with a distinguished $\ZZ_2$-grading
\begin{equation}V_\pm=\{v\in V:R(p)v=\pm v\}.
\end{equation}
Centrality of $p$ ensures that $R(g)$ is even with respect to this grading, for all $g\in \cG$. A $(\cG,p)$-equivariant Frobenius algebra is a Frobenius algebra $(A,m,\eta)$ with an action of $(\cG,p)$ that satisfies
\begin{equation}\label{equifrob1}
m(R(g)a\otimes R(g)b)=R(g)m(a\otimes b)
\end{equation}
and
\begin{equation}\label{equifrob2}
\eta(R(g)a,R(g)b)=\eta(a,b)
\end{equation}
for all $a,b\in A, g\in G$. As was true for the special case $\cG=\ZZ_2$, there are two notions of tensor product of these algebras: the usual one that forgets the distinguished $\ZZ_2$-grading and a supertensor product \eqref{supertensor} that remembers it. In both cases, the symmetry acts on the product as
\begin{equation}\label{stacksym}
R(g)(a_1\otimes a_2)=R_1(g)a_1\otimes R_2(g)a_2
\end{equation}
which is a special case of the rule
\begin{equation}\label{stackend}(\phi_1\otimes\phi_2)(a_1\otimes a_2)=(-1)^{|\phi_2||a_1|}\phi_1(a_1)\otimes\phi_2(a_2)
\end{equation}
for $\phi_1\otimes\phi_2\in\End(A_1)\otimes\End(A_2)$, where we have taken $R(g)=R_1(g)\otimes R_2(g)$.

We have argued in \cite{KTY} that bosonic phases with symmetry $G$ are classified by $G$-equivariant symmetric Frobenius algebras and that stacking of phases corresponds to the usual tensor product of their algebras. Here we will argue the fermionic analog: $(\cG,p)$-equivariant symmetric Frobenius algebras classify fermionic phases with symmetry $(\cG,p)$, for which stacking is governed by the supertensor product. In this language, bosonization means taking a $(\cG,p)$-equivariant algebra to a $\cG$-equivariant algebra by forgetting the distinguished involution $p$. Generically, if $\cG$ has more than one central involution, this map is many-to-one.


An equivariant module over a $(\cG,p)$-equivariant algebra $A$ is vector space $V$ with compatible actions of $A$ and $(\cG,p)$; that is, for every $a\in A$, we have a linear map $T(a)\in\End(V)$ such that $T(a)T(b)=T(ab)$, and for every $g\in G$, a linear map $\cQ(g)$ such that $\cQ(g)\cQ(h)=\cQ(gh)$. The compatibility condition reads
\begin{equation}\label{equimod}
T(R(g)a)=\cQ(g)T(a)\cQ(g)^{-1}
\end{equation}
Note that $T$ automatically respects the $\ZZ_2$-grading.

For a review of the classification of equivariant algebras and modules, we refer the reader to the prequel \cite{KTY}, which compiles some algebraic facts from \cite{Ostrik,Etingof}. There are two classes of algebras that will be especially useful in the present context, as they describe fermionic SRE phases. One class of algebras is those of the form $\End(U)$ for a projective representation $U$ of $\cG$. Each pair $(Q,U)$ has an associated class $[\omega]\in H^2(\cG,U(1))$ that measures the failure of $Q$ to be a homomorphism:
\begin{equation}\label{projrep}
Q(g)Q(h)=\exp(2\pi i\omega(g,h))Q(gh).
\end{equation} 
Each $[\omega]$ defines a Morita class of algebras and therefore a phase. Equivariant modules over $\End(U)$ are all of the form $U\otimes W$, where $W$ is a projective representation with class $-[\omega]$. When $\cG$ can be written as $G_b\times\{1,p\}$ for some group $G_b$ of bosonic symmetries, another class of equivariant algebras is those of the form $\End(U_b)\otimes \Cl(1)$ for a projective representation $(U_b,Q_b)$ of $G_b$. The group $G_b$ acts by conjugation on $\End(U_b)$. It also acts on the generator of $\Cl(1)$ by
\begin{equation}\label{betaaction}
\bar g: \Gamma\mapsto (-1)^{\beta(\bar g)}\Gamma,
\end{equation}
where $\beta:G_b\ra\ZZ_2$ is a homomorphism. Up to Morita-equivalence, algebras of this type depend only on the 1-cocycle $\beta$ and the 2-cocycle $\alpha$ on $G_b$ corresponding to the projective representation $Q_b$. While the bosonic phases built from these algebras have a broken $\ZZ_2$, their fermionic duals are nonetheless SRE phases.



\subsection{Equivariant fermionic MPS}

Let $(\cG,p)$ be a supergroup acting on the physical space $A$ by a unitary representation $R$. A $(\cG,p)$-invariant MPS tensor is a map $T:A\mapsto\End(U)$ such that $T(a)T(b)=T(ab)$ and
\begin{equation}T(R(g)a)=Q(g)T(a)Q(g)^{-1}
\end{equation}where the linear maps $Q(g)\in\End(U)$ form a projective representation of $(\cG,p)$ on $U$. For $X\in\End(U)$ satisfying the supercommutation rule \eqref{evenobs} or \eqref{oddobs}, the conjugate generalized MPS is
\begin{equation}\bra{\psi_T^X}=\Tr_U[XT(e_{i_1})\cdots T(e_{i_N})]\bra{i_1\cdots i_N}
\end{equation}in the NS sector and
\begin{equation}\bra{\psi_T^X}=\Tr_U[PXT(e_{i_1})\cdots T(e_{i_N})]\bra{i_1\cdots i_N}
\end{equation}in the R sector, where $P$ denotes $Q(p)$. More generally, we can insert $Q(g)$ instead of $P$:
\begin{equation}\label{twistedMPS}\bra{\psi_T^X}=\Tr_U[Q(g)XT(e_{i_1})\cdots T(e_{i_N})]\bra{i_1\cdots i_N}
\end{equation}
These are twisted sector states. When $\cG=G_b\times\{1,p\}$, states with twist $Q(\bar g,1)$ correspond to NS spin structure on a circle and a $G_b$ gauge field of holonomy $\bar g$, while states with twist $Q(\bar g,p)$ correspond to the Ramond spin structure on a circle and a $G_b$  gauge field of holonomy $\bar g$. When $\cG$ is non-split, one does not have spin structures and gauge fields, but a $\cG$-Spin structure, as discussed in Section \ref{equistatesum}.

Note that $\End(U)$ carries a genuine (not projective) action of $(\cG,p)$. By arguing as in \eqref{sameparity}, one can show that $\bra{\psi_T^X}$ transforms under $(\cG,p)$ in the same way as $X$.

\subsection{Fermionic SRE phases and their group structure}\label{SRE}


In this section, we restrict our attention to fermionic SRE phases, i.e. topological fermionic phases that are invertible under the stacking operation. These phases form a group under stacking. According to \cite{fidkit}, if the symmetry group $\cG$ splits as $G_b\times\ZZ_2$, each fermionic SRE phase corresponds to an element of the set
\begin{equation}\label{sptclassif}(\alpha,\beta,\gamma)\in H^2(G_b,U(1))\times H^1(G_b,\ZZ_2)\times\ZZ_2 . 
\end{equation}
If $G_b=\{1\}$, the two elements $(0,0,0)$ and $(0,0,1)$ correspond to the trivial and nontrivial Majorana chains, respectively. More generally, elements of the form $(\alpha,\beta,0)$ correspond to fermionic SRE phases that remain invertible after bosonization, while the bosonic duals of the fermionic SREs $(\alpha,\beta,1)$ are not SREs (they have a spontaneously broken $\ZZ_2$ but unbroken $G_b$).


If $\cG$ does not split, we claim that fermionic SRE phases are classified by pairs $(\alpha,\beta)$, where $\beta\in H^1(G_b,\ZZ_2)$, and $\alpha$ is a 2-cochain on $G_b$ with values in $U(1)$ satisfying $\delta\alpha=\frac12\rho\cup\beta$, i.e. for $\bar g,\bar h,\bar k,\in G_b$,
\begin{equation}\label{cup}\alpha(\bar g,\bar h)+\alpha(\bar{gh},\bar k)=\alpha(\bar h,\bar k)+\alpha(\bar g,\bar{hk})+\frac12\rho(\bar g,\bar h)\beta(\bar k)
\end{equation}
Here $\rho$ is the 2-cocycle on $G_b$ which encodes the multiplication in $\cG$. Certain pairs $(\alpha,\beta)$ correspond to equivalent SRE phases. Namely, adding to $\alpha$ an exact 2-cochain gives an equivalent SRE. Also, if we add to the 2-cocycle $\rho$ a coboundary of a 1-cochain $\mu$, $\alpha$ is shifted by $\frac12\mu\cup\beta$.

This classification can be understood from the standpoint of bosonization. Recall that $\cG$-invariant bosonic SREs are classified by group cohomology classes $[\omega]\in H^2(\cG,U(1))$ and arise from algebras of the form $A=\End(U)$ where $U$ is a projective representation of class $[\omega]$. Unlike the linear maps $R(g)$ of a genuine representation, the $Q(g)$ can be either even or odd with respect to $P:=Q(p)$. Using \eqref{projrep} and the centrality of $p$, it can be shown that $Q(g)$ and $Q(gp)$ have the same parity $\omega(p,g)-\omega(g,p)$; thus, one can define $\beta(\bar g):=|Q(g)|$. The function $\beta$ is clearly a homomorphism, and so defines a $\ZZ_2$-valued group $1$-cocycle of $G_b$. Given a trivialization $t$, one can re-express $\omega$ in terms of $\beta$ and a $U(1)$-valued group 2-cochain $\alpha$ of $G_b$ satisfying $\delta\alpha=\frac12\rho\cup\beta$ as follows:\footnote{When the extension splits, both $\alpha$ and $\beta$ are cocycles, and their equivalence to $\omega$ can be seen from the K\"unneth theorem for homology and the fact that $H^2(\cG,U(1))$ is the Pontryagin dual of $H_2(\cG,\ZZ)$.}
\begin{equation}\label{wat}\omega(g,h)=\alpha(\bar g,\bar h)+\frac12t(g)\beta(\bar h).
\end{equation}
Using \eqref{ctriv}, one can verify that $\eqref{cup}$ is equivalent to the cocycle condition for $\omega$. We prove in Appendix \ref{isom} that \eqref{wat} defines an isomorphism between $H^2(\cG,U(1))$ and the set of pairs $(\alpha,\beta)$, up to coboundaries.





When $\cG$ does not split, it is impossible to break $\ZZ_2$ without breaking $G_b=\cG/\ZZ_2$, so all fermionic SRE phases arise as fermionized bosonic SRE phases. Then the analysis above agrees with the result of \cite{fidkit} that, in the non-split case, fermionic SREs are classified by elements of $H^2(\cG,U(1))$ (modulo identifications).

But when $\cG$ splits, it is possible to break 
$\cG$ and still get an invertible fermionic phase. One can break $\cG$ down to any subgroup $H$ such that the quotient $\cG/H$ is a $\ZZ_2$ generated by $p$. Any such subgroup takes the form $H_\beta=\{ g\in\cG : t(g)=\beta(\bar g) \} $ for some homomorphism $\beta: G_b \rightarrow \ZZ_2$, and all homomorphisms give such a subgroup. This gives rise to a second class of fermionic SPTs - those whose bosonic duals are not invertible.

The algebras corresponding to these phases are of the form $A=\End(U_\beta)\otimes \Cl(1)$ for some projective representation $(U_\beta,Q_\beta)$ of $H_\beta$. Let $h\in H_\beta$, $M\in\End(U_\beta)$, $m\in\ZZ_2$. The subgroup and quotient act on $A$ as
\begin{equation}R(h):M\otimes\Gamma^m\mapsto Q_\beta(h)^{-1}MQ_\beta(h)\otimes\Gamma^m,
\end{equation}
\begin{equation}R(p):M\otimes\Gamma^m\mapsto(-1)^mM\otimes\Gamma^m
\end{equation}
This action is a special case of the more general rule discussed in Section 4.3 of \cite{KTY}. In terms of $\cG$,
\begin{equation} R(g)=R(\bar g,\beta(\bar g))\cdot R(p)^{t(g)+\beta(\bar g)}:M\otimes\Gamma^m\mapsto (-1)^{m(t(g)+\beta(\bar g))}Q_\beta(\bar g,\beta(\bar g))^{-1}MQ_\beta(\bar g,\beta(\bar g))\otimes\Gamma^m \label{g1alg}
\end{equation}
as claimed in \eqref{betaaction} (after setting $t(\bar g,1)=0$). Note that $\beta$, which encodes the action of the symmetry on fermions, can be offset by changing the trivialization $t$, i.e. the splitting isomorphism $\cG\xrightarrow{\sim}G_b\times\ZZ_2$. As a projective representation, $Q_\beta$ is characterized by a class $[\alpha]\in H^2(H,U(1))\simeq H^2(G_b,U(1))$.



We have shown that $(\cG,p)$-equivariant fermionic SRE phases can be characterized by pairs $(\alpha,\beta)$ and - if $\cG$ is split - an additional $\ZZ_2$ label $\gamma$ that represents a $C\ell(1)$ factor in the algebra. This parameterization is useful for discussing stacking of fermionic phases, which is different from the standard group structure on  $H^2(\cG,U(1))$ (the latter describes bosonic stacking). First, since $\Cl(1)\fotimes\Cl(1)\simeq\Cl(2)$ is Morita-equivalent to $\CC$, the $\gamma$ parameters must simply add up under stacking. Second, if we consider two phases with parameters $(\alpha_1,\beta_1,0)$ and $(\alpha_2,\beta_2,0)$ corresponding to two $\cG$-equivariant algebras $(Q_1,U_1)$ and $(Q_2,U_2)$, the supertensor product is a $\cG$-equivariant algebra $(Q,U)$, where $U=U_1\fotimes U_2$ and $Q=Q_1\fotimes Q_2$. We can easily compute: 
\begin{align}Q(g)Q(h)&=(Q_1(g)\fotimes Q_2(g))(Q_1(h)\fotimes Q_2(h))\nonumber\\&=(-1)^{\beta_2(\bar g)\beta_1(\bar h)}Q_1(g)Q_1(h)\fotimes Q_2(g)Q_2(h)\nonumber\\&=(-1)^{\beta_2(\bar g)\beta_1(\bar h)}\exp(2\pi i\alpha_1(\bar g,\bar h))(-1)^{t(g)\beta_1(\bar h)}\exp(2\pi i\alpha_2(\bar g,\bar h))(-1)^{t(g)\beta_2(\bar h)}Q_1(gh)\fotimes Q_2(gh)\nonumber\\&=\exp(2\pi i(\alpha_1+\alpha_2+\frac12\beta_2\cup\beta_1))(\bar g,\bar h)(-1)^{t(g)(\beta_1+\beta_2)(\bar h)}Q(gh).
\end{align}
Thus the group structure in this case is
$$
(\alpha_1,\beta_1,0)+(\alpha_2,\beta_2,0)=(\alpha_1+\alpha_2+\frac12\beta_1\cup\beta_2,\beta_1+\beta_2,0).
$$
Note that $\beta_1\cup\beta_2$ differs from $\beta_2\cup\beta_1$ by an exact term, and thus the difference between them is inessential.
Based on these two special cases it is easy to guess that the group structure induced by stacking is
\begin{equation}\label{groupstructure}
(\alpha_1,\beta_1,\gamma_1)+(\alpha_2,\beta_2,\gamma_2)=(\alpha_1+\alpha_2+\frac12\beta_1\cup\beta_2,\beta_1+\beta_2,\gamma_1+\gamma_2).
\end{equation}
 This is verified in Appendix \ref{sec:grouplaw}

The set of triples $(\alpha,\beta,\gamma)$ with this group law is isomorphic to the spin-cobordism group $\Omega^2_{Spin}(BG_b)$ \cite{GK}. This agrees with the proposal of \cite{KTTW} about the classification of fermionic SRE phases. In the non-split case, the group structure is given by the same formulas, except that $\gamma$ is set to zero, and $\alpha$ is not closed, but satisfies the equation $\delta\alpha=\frac12\rho\cup\beta$.

If $\cG$ splits, the isomorphism $\cG\simeq G_b\times\ZZ_2$ may be taken as part of the physical data. This means that one fixes the action of $G_b$ on fermions as well as on bosons. Alternatively, if one regards this isomorphism as unphysical, one only fixes the action of $G_b$ on bosons, while the action on fermions is fixed only up certain signs. So far we have been taking the former viewpoint. If we take the latter viewpoint, we also need to understand how the parameters $(\alpha,\beta,\gamma)$ change when we change the action of $G_b$ on fermions. Given a particular action of ${\bar g}\in G_b$, any other action which acts in the same way on bosons differs from it by $p^{\mu(\bar g)}$, where $p$ is fermion parity and $\mu:G_b\ra\ZZ_2$ is a homomorphism. If we define ${\tilde Q}(\bar g)= Q(\bar g) P^{\mu(\bar g)}$, we have
$$
{\tilde Q}(\bar g){\tilde Q}(\bar h)=\exp(2\pi i \alpha(\bar g,\bar h))(-1)^{\mu(\bar g)\beta(\bar h)} {\tilde Q}(\bar g\bar h),
$$
and 
$$
P{\tilde Q}(\bar g) P^{-1}=(-1)^{\beta(\bar g)} {\tilde Q}(\bar g).
$$
This implies that for $\gamma=0$ the parameter $\beta$ is unchanged, while $\alpha\mapsto\alpha+\frac12 \mu\cup\beta$. For $\gamma=1$ the situation is different, since fermion parity acts trivially on $U$, and thus $\alpha$ is not modified. But it acts nontrivially on the generator of $\Cl(1)$, so that the new $G_b$ transformation multiplies it by $(-1)^{\beta(\bar g)+\mu(\bar g)}$. Thus $\beta\mapsto\beta+\mu$. Thus if we do not fix the action of $G_b$ on fermions, all fermionic SRE phases  with $\gamma=1$ and a fixed $[\alpha]$ are equivalent. This agrees with \cite{fidkit}.

\subsection{Two examples with $G_b=\ZZ_2$}

Let us consider the case $G_b=\ZZ_2=\{1,b\}$. There are two extensions of $G_b$ by fermionic parity $\ZZ_2^\cF=\{1,p\}$: one is $\ZZ_2\times\ZZ_2=\ZZ[b]/(b^2)\times\ZZ[p]/(p^2)$; the other is $\ZZ_4=\ZZ[b,p]/(b^2-p)$.

First take $\cG=\ZZ_2\times\ZZ_2$. Consider algebras of the form $A=\End(U)$, where $U$ is a projective representation of $\cG$. Each is characterized by a class $[\omega]\in H^2(\ZZ_2\times\ZZ_2,U(1))=\ZZ_2$. The two options for $[\omega]$ have cocycle representatives
\begin{equation}\omega_0(g,h)=0\quad\text{ and }\quad\omega_1(g,h)=\frac12 g_2h_1
\end{equation}where $g=(g_1,g_2)$, $h=(h_1,h_2)$. On the bosonic side of the duality, we think of $\omega_0$ as describing the trivial phase and $\omega_1$ as describing a nontrivial SRE. Alternatively, one can replace each $\omega$ by a pair $(\alpha,\beta)$. There is only the trivial $[\alpha]\in H^2(\ZZ_2,U(1))$. There are two $\beta$'s: $\beta_0(b)=0$ and $\beta_1(b)=1$. These correspond to $\omega_0$ and $\omega_1$, respectively, as
\begin{equation}\omega_i(g,h)=\frac12 t(g)\beta_i(b(g))
\end{equation}where $b(g)=g_1$ and $t(g)=g_2$. On the fermionic side, $\beta_0$ describes a trivial phase and $\beta_1$ a nontrivial SRE.

Now consider breaking the symmetry down to any of the three $\ZZ_2$ subgroups of $\cG$; this means considering algebras $A=\text{Ind}_H^\cG(\End(U))$ for projective representations $U$ of the unbroken $H=\ZZ_2$. Since $H^2(\ZZ_2,U(1))$ is trivial, the only possibility (up to Morita equivalence) is $A=C\ell(1)$, graded by $\cG/H$. On the bosonic side, each choice of $H$ is a different non-invertible phase. As fermionic phases, the $G_b$-graded $C\ell(1)$ is a symmetry-broken phase, while the $\ZZ_2^\cF$-graded $C\ell(1)$ is a nontrivial Majorana-chain phase $(0,\beta_0,1)$. Breaking down to the diagonal $\ZZ_2$ gives a $p$-graded $C\ell(1)$ on which the bosonic symmetry acts non-trivially, i.e. $(0,\beta_1,1)$.

Now take $\cG=\ZZ_4$. The extension class is represented by the $2$-cocycle $\rho(b,b)=1$. There is only the trivial class $[\omega]\in H^2(\ZZ_4,U(1))=\{1\}$. Meanwhile, there are two $\beta$'s: $\beta_0$ and $\beta_1$ as before. They satisfy $\rho\cup\beta_0=0$ and $\rho\cup\beta_1(b,b,b)=1$. The trivial $\alpha$ is the unique solution to $\delta\alpha=\rho\cup\beta_0$, and one can show that there are no solutions to $\delta\alpha=\rho\cup\beta_1$. In summary, there is only one pair $(\alpha,\beta)$ - it's the trivial one.

Consider breaking the only subgroup $\ZZ_2^\cF$. The corresponding algebra is the $G_b$-graded $C\ell(1)$, which, as before, describes a symmetry-broken phase in both the bosonic and fermionic pictures.

\begin{figure}[h]\label{Z2classif}
\centering
\begin{subfigure}{0.5\textwidth}
\centering
\begin{tabular}{ c | c | c | c }
bosonic & $(H,\omega)$ & $(\alpha,\beta,\gamma)$ & fermionic \\
 \hline			
 trivial & $(\cG,\omega_0)$ & $(0,\beta_0,0)$ & trivial \\
 BSRE & $(\cG,\omega_1)$ & $(0,\beta_1,0)$ & FSRE \\
 SB & $(\ZZ_2^\cF,1)$ & n/a & SB \\
  SB & $(G_b,1)$ & $(0,\beta_0,1)$ & FSRE \\
  SB & $(\langle bp\rangle,1)$ & $(0,\beta_1,1)$ & FSRE \\
  \hline  
\end{tabular}
\caption{Phases with $\cG=\ZZ_2\times\ZZ_2$}
\end{subfigure}%
\begin{subfigure}{0.5\textwidth}
\centering
\begin{tabular}{ c | c | c | c }
bosonic & $(H,\omega)$ & $(\alpha,\beta)$ & fermionic \\
  \hline			
  trivial & $(\cG,\omega_0)$ & $(0,\beta_0)$ & trivial \\
  SB & $(G_b,1)$ & n/a & SB \\
  \hline  
\end{tabular}
\caption{Phases with $\cG=\ZZ_4$}
\end{subfigure}
\caption{Phase classification for the $G_b=\ZZ_2$ symmetry groups}
\end{figure}

\subsection{State-sum for the equivariant fermionic theory}\label{equistatesum}

In Section \ref{MPSfromTQFT}, we observed that fermionic MPS arise from the state-sum for a spin-TQFT evaluated on an annulus diagram. A similar story can be told about equivariant fermionic MPS. Now we will define a state-sum for equivariant spin-TQFTs and recover the MPS \eqref{twistedMPS} as states on an annulus.

We will focus on the case where the total symmetry group $\cG$ splits as a product of $G_b$ and $\ZZ_2$ and then indicate the modifications needed in the non-split case. A $G_b$-equivariant spin-TQFT is defined in the same way as an ordinary spin TQFT, except that spin manifolds are replaced with spin manifolds equipped with principal $G_b$-bundles. Since $G_b$ is finite, a $G_b$-principal bundle is completely characterized by its holonomies on non-contractible cycles. We will denote by $\cA$ the collection of all holonomies. When working on manifolds with boundaries, it is convenient to fix a marked point and a trivialization of the bundle at the this point on each boundary, so that the holonomy around each of these circles is a well-defined element of $G_b$ rather than a conjugacy class.



The algebraic input for the state-sum construction is $G_b\times\ZZ_2$-equivariant semisimple Frobenius algebra $A$. The geometric data are a closed oriented two-dimensional manifold $\Sigma$ equipped with a $G_b$-bundle and a spin structure. To define the state-sum, we also choose a marked skeleton $\Gamma$, then a trivialized $G_b$-bundle can be represented as a decoration of each oriented edge with an element $g\in G_b$. Reversing an edge orientation replaces $g$ with $g^{-1}$. We impose a flatness condition: the product of group labels around the boundary of each 2-cell is the identity. Equivalently, we can use the dual triangulation $\Gamma^*$: each dual edge is labeled by a group element, and the flatness condition says that the cyclically-ordered product of group elements on dual edges meeting at each dual vertex is the identity. One can think of the dual edges as domain walls and the dual edge labels as the $G_b$ transformations due to moving across them. 

The state-sum is defined as follows. Given a skeleton with a principal bundle, color the edges with pairs of elements $e_i$ of some homogeneous basis of $A$. The weight of a coloring is the product of structure constants $C_{ijk}$ over vertices (with indices cyclically ordered by orientation) and terms $R(g)^{i}{}_k\eta^{kj}$ over edges times the spin-dependent Koszul sign $\sigma_s$. The partition sum is the sum of the weights over colorings; the holonomies $\cA$, which represent a background gauge field, are not summed over.




To incorporate brane boundaries, choose a $G_b\times\ZZ_2$-equivariant $A$-module $U$ for each boundary component. Color the boundary edges by pairs of elements $f^U_\mu$ of a homogeneous basis of $U$ - one for each vertex sharing the edge. The weight of a coloring is the usual weight times a factor of $T^{i\mu}{}_\nu$ for each boundary vertex and $\cQ(g)^\mu{}_\nu$ for each boundary edge.

As in the non-equivariant case, the partition sum is a spin-topological invariant. It also does not depend on the choice of trivialization of the principal bundle; in other words, it is gauge invariant. Invariance is ensured by the equivariance conditions \eqref{equifrob1}, \eqref{equifrob2}, and \eqref{equimod}. In fact, one can evaluate the partition function in a closed form when the boundary is empty. Let $A=\End(U)\otimes \Cl(1)$ for some projective representation of $G_b$ with a 2-cocycle $\alpha$, and the action of $G_b$ on $\Cl(1)$ determined by a homomorpism $\beta:G_b\ra\ZZ_2$. 
It is easy to see that the partition function factorizes into a product of the partition function corresponding to $\End(U)$ and the partition function corresponding to $\Cl(1)$. The former factor is the partition function of a bosonic SRE phases, i.e. $\exp(2\pi i\int_\Sigma\alpha)$ \cite{KTY}. The latter one is essentially the Arf invariant, modified by additional signs from the edges $e$ for which $\beta(e)=1$:
\begin{equation}
2^{-b_1(\Sigma)/2} \sum_{[a]\in H_1(\Sigma,\ZZ_2)} \sigma_s(a) (-1)^{\sum_{e\in a}\beta(\cA(e))}.
\end{equation}
Using the property (\ref{sigmasvariation}), the definition of the Arf invariant, and the identity $\Arf(s+a)=\Arf(s) \sigma_s(a)$ \cite{Atiyah}, we can write this as
\begin{equation}
\Arf(s+\beta(\cA))=\Arf(s)\sigma_s(\beta(\cA)).
\end{equation}
Thus partition function of the fermionic SRE with the parameters $(\alpha,\beta,1)$ is
\begin{equation}
\exp(2\pi i\int_\Sigma\alpha) \sigma_s(\beta(\cA))\Arf(s).
\end{equation}
Tensoring with another copy of $\Cl(1)$ multiplies this by another factor $\Arf(s)$, so that the partition function of the fermionic SRE with the parameters $(\alpha,\beta,0)$ is
\begin{equation}
\exp(2\pi i\int_\Sigma\alpha) \sigma_s(\beta(\cA)).
\end{equation}

We can also recover the equivariant MPS wavefunctions from the state sum. First suppose $A=\End(U)$, i.e. the parameter $\gamma=0$. An equivariant module over $A$ is of the form $M=U\otimes W$, where $(U,Q)$ and $(W,S)$ have projective actions of $\cG$ characterized by opposite cocycles. Consider the annulus where one boundary is a brane boundary labeled by $M$ and the other is a cut boundary. We work with a skeleton on the annulus such that each boundary is divided into $N$ intervals, and let $g_{i,i+1}$ denote the group label between vertices $i$ and $i+1$. A computation similar to that of Section \ref{MPSfromTQFT} gives the state
\begin{equation}\bra{\psi_T}=\sum\Tr_{U\otimes W}[T(e_{i_1})\cQ(g_{12})\cdots T(e_{i_N})\cQ(g_{N1})]\bra{i_1\cdots i_N}
\end{equation}
which, after performing gauge transformations and LU transformations, can be put in the form
\begin{equation}
\bra{\psi_T}=\sum\Tr_{U\otimes W}[\cQ(g)T(e_{i_1})\cdots T(e_{i_N})]\bra{i_1\cdots i_N}
\end{equation}where $g=g_{12}\cdots g_{N1}$. 
Since $\cQ=Q\otimes S$ and $T(e_i)$ has the form $T(e_i)\otimes\mathds{1}_W$, the trace factorizes:
\begin{equation}\bra{\psi_T}=\Tr_W[S(g)]\sum\Tr_U[Q(g)T(e_{i_1})\cdots T(e_{i_N})]\bra{i_1\cdots i_N}.
\end{equation}
Up to normalization, this is the MPS \eqref{twistedMPS}.

The case $A=\End(U_\beta)\otimes C\ell(1)$ is similar. An indecomposable module over $A$ is of the form $U\otimes W\otimes V$, where $U$ and $W$ carry projective $H_\beta$ actions of opposite cocycles and $V=\CC^{1|1}$ is the $C\ell(1)$-module considered in Section \ref{nontriv}. The action of $\cG$ is determined by $\cQ(h)=Q_\beta(h)\otimes S(h)\otimes\mathds{1}$ and $\cQ(p)(M\otimes u_\pm)=\pm M\otimes u_\pm$. The argument proceeds as before, with the trace over $W$ factoring out. We are left with an expression of the form \eqref{twistedMPS} where the trace is over $U\otimes V$, the most general indecomposable MPS tensor over $A$.




Let us now discuss the non-split case. If $\cG$ is a nontrivial extension of $G_b$ by fermion parity, it is no longer true that a $\cG$-equivariant algebra defines a $G_b$-equivariant spin-TQFT. Rather, it defines a $\cG$-Spin TQFT \cite{KTTW}. A $\cG$-Spin structure on a manifold $X$ is a $G_b$ gauge field $\cA$ on $X$  together with a trivialization of the $\ZZ_2$-valued 2-cocycle $w_2-\rho(\cA)$, where $\rho(\cA)$ is the pull-back of $\rho$ from $BG_b$ to $X$ and $w_2$ is a 2-cocycle representing the $2^\text{nd}$ Stiefel-Whitney class of $X$. 
Now, if $X$ is a Riemann surface $\Sigma$, $[w_2]$ is always zero, so $[\rho(\cA)]$ must be trivial too. Instead of choosing a trivialization of $w_2-\rho(\cA)$, we can choose a trivialization $s$ of $w_2$ and a trivialization $\tau$ of $\rho(\cA)$. That is, we choose $\ZZ_2$-valued 1-cochains $s$ and $\tau$ such that $\delta s=w_2$ and $\delta\tau=\rho(\cA)$. These data are redundant: we can shift both $s$ and $\tau$ by $\psi\in H^1(\Sigma,\ZZ_2)$.

We can now proceed as in the split case. Instead of a triple $(\alpha,\beta,\gamma)$ we have a pair $(\alpha,\beta)$ where $\beta\in H^1(G_b,\ZZ_2)$ and $\alpha$ is a 2-cochain on $G_b$ with values in $U(1)$ satisfying $\delta\alpha=\frac12\rho\cup\beta$. These data parameterize a 2-cocycle on $\cG$. As shown above, the pairs $(\alpha,\beta)$ and $(\alpha+\frac12 \mu\cup\beta,\beta)$ correspond to the same 2-cocycle on $\cG$, for any $\mu\in H^1(G_b,\ZZ_2)$. The partition function is evaluated exactly in the same way as in the split case, except that $\alpha$ is no longer closed, and an extra correction factor is needed to ensure the invariance of the partition function under a change of triangulation or a $G_b$ gauge transformation. This correction factor is
\begin{equation}\label{sigmafactor}
(-1)^{\int_\Sigma \tau\cup\beta(\cA)}
\end{equation}
where $\tau$ is a trivialization of $\rho(\cA)$ which is part of the definition of the $\cG$-Spin structure on $\Sigma$. Thus the partition function is
\begin{equation}
\exp(2\pi i\int_\Sigma\alpha(\cA))(-1)^{\int_\Sigma \tau\cup\beta(\cA)}\sigma_s(\beta(\cA)).
\end{equation}
Using (\ref{sigmasvariation}) one can easily see that the partition function is invariant under shifting both $\tau$ and $s$ by any $\psi\in H^1(\Sigma,\ZZ_2)$. One can also see that the partition function is invariant under shifting $\alpha$ by $\frac12 \mu\cup\beta$ for any $\mu\in H^1(G_b,\ZZ_2)$ if we simultaneously shift $\tau\mapsto \tau+\mu(\cA)$. 

Returning to the split case, we can examine the effect of treating the isomorphism $\cG\simeq G_b\times\ZZ_2$ as unphysical. Every two such isomorphisms differ by a homomorphism $\mu:G_b\ra\ZZ_2$. The effect this has on the data $(\alpha,\beta,\gamma)$ has been described in section \ref{SRE}:
\begin{equation}
\alpha\mapsto\alpha+(1-\gamma)\frac12\mu\cup\beta,\qquad \beta\mapsto\beta+\gamma\mu,\qquad\gamma\mapsto \gamma.
\end{equation}
Using the properties of $\sigma_s$ and the Arf invariant, it is easy to check that the partition function is unaffected by these substitutions if we simultaneously shift the spin structure:
\begin{equation}
s\mapsto s+\mu(\cA).
\end{equation}
This can be interpreted as a special case of an equivalence relation between different spin structures which define the same $\cG$-Spin structure.

\appendix

\section{Diagrams for the ground states}\label{sec:diagrams}

These diagrams are used in the argument of Section \ref{hamsection}.

\begin{figure}[H]
$C_2(\psi_2)=$
\begin{tikzpicture}[baseline={([yshift=-.5ex]current bounding box.center)}]
\centering

\path (0,0) (1,0) node[circle,inner sep=1pt, draw](X){$X$};

\draw(0,0)--(X);
\draw(X)--(3,0);
\draw(2,0)--(2,-1);
\draw[magenta,line width=0.5mm](1,-1)--(4,-1);
\draw(3,-1)--(3,-2);

\end{tikzpicture} $=$\hspace{1cm}\begin{tikzpicture}[baseline={([yshift=-.5ex]current bounding box.center)}]

\centering

\path (0,0) (1,0) node[circle,inner sep=1pt, draw](X){$X$};

\draw(0,0)--(X);
\draw(X)--(2,-0.5);
\draw(3,-0.5)--(4,0)--(5,0);
\draw[magenta, line width=0.5mm] (2,-0.5)--(3,-0.5);
\draw[magenta,line width=0.5mm] (2,-0.5)--(1,-1)--(0,-1);
\draw [magenta,line width=0.5mm](3,-0.5)--(4,-1)--(5,-1);
\draw (4.5,-1)--(4.5,-2);

\end{tikzpicture} \hspace{0.5cm} $=$ \begin{tikzpicture}[baseline={([yshift=-.5ex]current bounding box.center)}]

\centering

\path (2,0.5) node[circle,inner sep=1pt, draw](X){$X$};

\draw[magenta,line width=0.5mm] (0,0) -- (1,0);
\draw(1,0)--(1.5,0.5)--(X);
\draw (X)--(2.5,0.5)--(3,0);
\draw[magenta,line width=0.5mm](3,0)--(4,0);
\draw[magenta,line width=0.5mm](1,0)--(2,-0.5)--(3,0);
\draw(2,-0.5)--(2,-1.5);
\end{tikzpicture} \hspace{0.5cm}$= \eta_X$ \hspace{1cm} \begin{tikzpicture}[baseline={([yshift=-.5ex]current bounding box.center)}]

\centering

\path (1,0) node[circle,inner sep=1pt, draw](X){$X$};
\draw[magenta,line width=0.5mm] (0,0) -- (X);
\draw[magenta,line width=0.5mm] (X)--(2,0);
\draw(2,0)--(3,0);
\draw[magenta,line width=0.5mm](3,0)--(4,0);
\draw[magenta,line width=0.5mm](2,0)--(2.5,-1)--(3,0);
\draw (2.5,-1)--(2.5,-2);

 \end{tikzpicture} \hspace{0.5cm}$=\eta_X$\begin{tikzpicture}[baseline={([yshift=-.5ex]current bounding box.center)}]

\centering

\path (1,0) node[circle,inner sep=1pt, draw](X){$X$};
\draw[magenta,line width=0.5mm](0,0) -- (X)--(4,0);
\draw (3,0)--(3,-1);
\end{tikzpicture}

\vspace{0.5cm} $= \eta_X \psi_1$
\caption{Diagrammatic proof of $C_2 \langle \psi_1| = \eta_X \langle \psi_2|$. }
\label{fig:p2}
\end{figure}
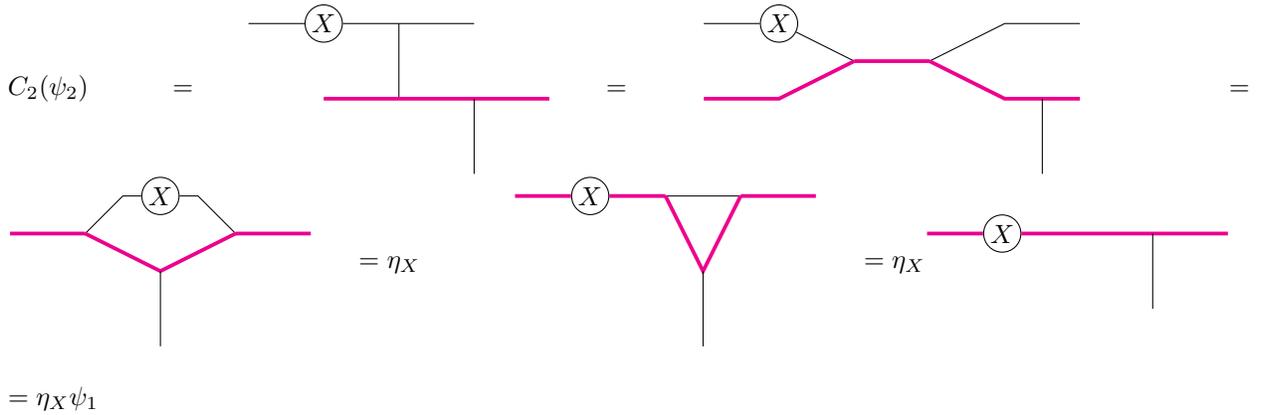

\begin{figure}[H]
$C_1(\psi_2)=$
\begin{tikzpicture}[baseline={([yshift=-.5ex]current bounding box.center)}]
\centering

\path (0,0) (1,0) node[circle,inner sep=1pt, draw](X){$X$};

\draw[magenta,line width=0.5mm](0,0)--(X);
\draw[magenta,line width=0.5mm](X)--(3,0);
\draw(2,0)--(2,-1);
\draw(1,-1)--(4,-1);
\draw(3,-1)--(3,-2);

\end{tikzpicture} $=$\hspace{1cm}\begin{tikzpicture}[baseline={([yshift=-.5ex]current bounding box.center)}]

\centering

\path (0,0) (1,0) node[circle,inner sep=1pt, draw](X){$X$};

\draw[magenta,line width=0.5mm] (0,0)--(X);
\draw[magenta,line width=0.5mm] (X)--(2,-0.5)--(3,-0.5)--(4,0)--(5,0);
\draw (2,-0.5)--(1,-1)--(0,-1);
\draw (3,-0.5)--(4,-1)--(5,-1);
\draw (4.5,-1)--(4.5,-2);

\end{tikzpicture} \hspace{0.5cm} $=$ \begin{tikzpicture}[baseline={([yshift=-.5ex]current bounding box.center)}]

\centering

\path (2,0.5) node[circle,inner sep=1pt, draw](X){$X$};

\draw[magenta,line width=0.5mm] (0,0) -- (1,0)--(1.5,0.5)--(X);
\draw[magenta,line width=0.5mm] (X)--(2.5,0.5)--(3,0)--(4,0);
\draw(1,0)--(2,-0.5)--(3,0);
\draw(2,-0.5)--(2,-1.5);
\end{tikzpicture}  \hspace{0.5cm} $=$ \hspace{1cm} \begin{tikzpicture}[baseline={([yshift=-.5ex]current bounding box.center)}]

\centering

\path (1,0) node[circle,inner sep=1pt, draw](X){$X$};
\draw [magenta,line width=0.5mm](0,0) -- (X);
\draw [magenta,line width=0.5mm](X)--(4,0);
\draw(2,0)--(2.5,-1)--(3,0);
\draw (2.5,-1)--(2.5,-2);

\end{tikzpicture}  \hspace{0.5cm}$=$\hspace{0.5cm}\begin{tikzpicture}[baseline={([yshift=-.5ex]current bounding box.center)}]

\centering

\path (1,0) node[circle,inner sep=1pt, draw](X){$X$};
\draw[magenta,line width=0.5mm](0,0) -- (X)--(4,0);
\draw (3,0)--(3,-1);
\end{tikzpicture}

\vspace{0.5cm} $= \psi_2$
\caption{Diagrammatic proof of $C_1 \langle \psi_2| = \langle \psi_2|$. }
\label{fig:p3}
\end{figure}

\begin{figure}[H]
$C_2(\psi_2)=$
\begin{tikzpicture}[baseline={([yshift=-.5ex]current bounding box.center)}]
\centering

\path (0,0) (1,0) node[circle,inner sep=1pt, draw](X){$X$};

\draw[magenta,line width=0.5mm](0,0)--(X);
\draw[magenta,line width=0.5mm](X)--(3,0);
\draw(2,0)--(2,-1);
\draw[magenta,line width=0.5mm](1,-1)--(4,-1);
\draw(3,-1)--(3,-2);

\end{tikzpicture} $=$\hspace{1cm}\begin{tikzpicture}[baseline={([yshift=-.5ex]current bounding box.center)}]

\centering

\path (0,0) (1,0) node[circle,inner sep=1pt, draw](X){$X$};

\draw[magenta,line width=0.5mm] (0,0)--(X);
\draw[magenta,line width=0.5mm] (X)--(2,-0.5);
\draw[magenta,line width=0.5mm] (3,-0.5)--(4,0)--(5,0);
\draw (2,-0.5)--(3,-0.5);
\draw[magenta,line width=0.5mm] (2,-0.5)--(1,-1)--(0,-1);
\draw [magenta,line width=0.5mm](3,-0.5)--(4,-1)--(5,-1);
\draw (4.5,-1)--(4.5,-2);

\end{tikzpicture} \hspace{1cm} $=$ \begin{tikzpicture}[baseline={([yshift=-.5ex]current bounding box.center)}]

\centering

\path (2,0.5) node[circle,inner sep=1pt, draw](X){$X$};

\draw (0,0) -- (1,0);
\draw[magenta,line width=0.5mm](1,0)--(1.5,0.5)--(X);
\draw[magenta,line width=0.5mm] (X)--(2.5,0.5)--(3,0);
\draw(3,0)--(4,0);
\draw[magenta,line width=0.5mm](1,0)--(2,-0.5)--(3,0);
\draw(2,-0.5)--(2,-1.5);
\end{tikzpicture} \hspace{0.5cm}$= \eta_X$ \hspace{1cm} \begin{tikzpicture}[baseline={([yshift=-.5ex]current bounding box.center)}]

\centering

\path (1,0) node[circle,inner sep=1pt, draw](X){$X$};
\draw (0,0) -- (X);
\draw (X)--(2,0);
\draw[magenta,line width=0.5mm](2,0)--(3,0);
\draw(3,0)--(4,0);
\draw[magenta,line width=0.5mm](2,0)--(2.5,-1)--(3,0);
\draw (2.5,-1)--(2.5,-2);

\end{tikzpicture} $=\eta_X$\hspace{1cm}\begin{tikzpicture}[baseline={([yshift=-.5ex]current bounding box.center)}]

\centering

\path (1,0) node[circle,inner sep=1pt, draw](X){$X$};
\draw(0,0) -- (X)--(4,0);
\draw (3,0)--(3,-1);
\end{tikzpicture}

\vspace{0.5cm} $= \eta_X \psi_1$
\caption{Diagrammatic proof of $C_2 \langle \psi_2| = \eta_X \langle \psi_1|$.}
\label{fig:p4}
\end{figure}

\begin{figure}[H]
$C_3(\psi_3)=$
\begin{tikzpicture}[baseline={([yshift=-.5ex]current bounding box.center)}]
\centering

\path (0,0) (1,0) node[circle,inner sep=1pt, draw](X){$X$};

\draw(0,0)--(X);
\draw[magenta,line width=0.5mm](X)--(2,0);
\draw(2,0)--(3,0);
\draw[magenta,line width=0.5mm](2,0)--(2,-1);
\draw(1,-1)--(4,-1);
\draw[magenta,line width=0.5mm](2,-1)--(3,-1)--(3,-2);

\end{tikzpicture} $=$\hspace{0.5cm}\begin{tikzpicture}[baseline={([yshift=-.5ex]current bounding box.center)}]

\centering

\path (0,0) (1,0) node[circle,inner sep=1pt, draw](X){$X$};

\draw (0,0)--(X);
\draw[magenta,line width=0.5mm] (X)--(2,-0.5)--(3,-0.5);
\draw (3,-0.5)--(4,0)--(5,0);
\draw (2,-0.5)--(1,-1)--(0,-1);
\draw [magenta,line width=0.5mm](3,-0.5)--(4,-1)--(4.5,-1);
\draw(4.5,-1)--(5,-1);
\draw[magenta,line width=0.5mm] (4.5,-1)--(4.5,-2);

\end{tikzpicture} \hspace{0.5cm} $=$ \begin{tikzpicture}[baseline={([yshift=-.5ex]current bounding box.center)}]

\centering

\path (2,0.5) node[circle,inner sep=1pt, draw](X){$X$};

\draw[magenta,line width=0.5mm] (0,0) -- (1,0);
\draw(1,0)--(1.5,0.5)--(X);
\draw[magenta,line width=0.5mm] (X)--(2.5,0.5)--(3,0);
\draw[magenta,line width=0.5mm](3,0)--(4,0);
\draw[magenta,line width=0.5mm](1,0)--(2,-0.5)--(2,-1.5);
\draw(2,-0.5)--(3,0);
\end{tikzpicture} \hspace{0.5cm}$= \eta_X$ \hspace{1cm} \begin{tikzpicture}[baseline={([yshift=-.5ex]current bounding box.center)}]

\centering

\path (1,0) node[circle,inner sep=1pt, draw](X){$X$};
\draw[magenta,line width=0.5mm] (0,0) -- (X);
\draw (X)--(2,0);
\draw[magenta,line width=0.5mm](2,0)--(3,0);
\draw[magenta,line width=0.5mm](3,0)--(4,0);
\draw[magenta,line width=0.5mm](2,0)--(2.5,-1);
\draw(2.5,-1)--(3,0);
\draw[magenta,line width=0.5mm] (2.5,-1)--(2.5,-2);

\end{tikzpicture} $=\eta_X$\hspace{1cm}\begin{tikzpicture}[baseline={([yshift=-.5ex]current bounding box.center)}]

\centering

\path (1,0) node[circle,inner sep=1pt, draw](X){$X$};
\draw[magenta,line width=0.5mm](0,0) -- (X);
\draw(X)--(3,0);
\draw[magenta,line width=0.5mm](3,0)--(4,0);
\draw[magenta,line width=0.5mm] (3,0)--(3,-1);
\end{tikzpicture}

\vspace{0.5cm} $= \eta_X \psi_4 = \psi_3$
\caption{Diagrammatic proof of $C_3 \langle \psi_3| =  \langle \psi_3|$.}
\label{fig:p5}
\end{figure}

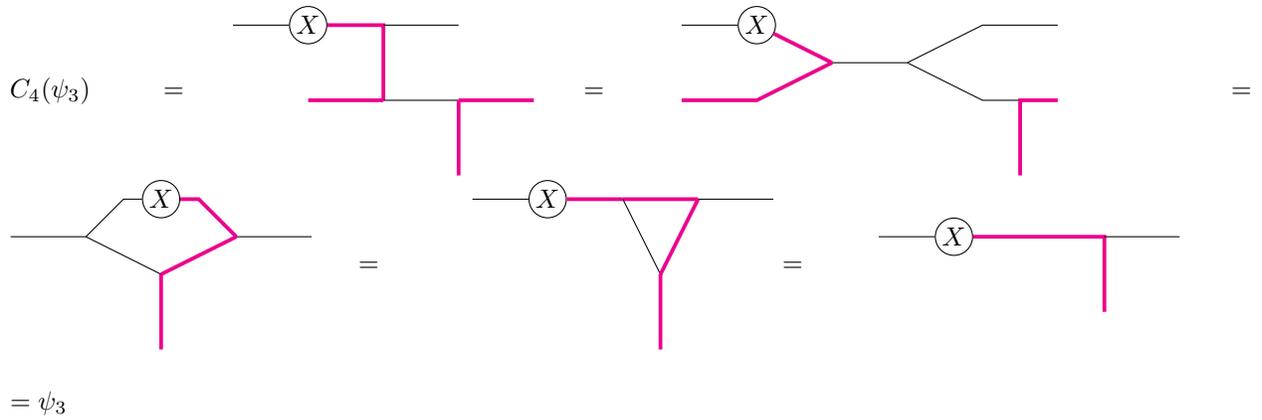
\begin{figure}[H]
$C_4(\psi_3)=$
\begin{tikzpicture}[baseline={([yshift=-.5ex]current bounding box.center)}]
\centering

\path (0,0) (1,0) node[circle,inner sep=1pt, draw](X){$X$};

\draw(0,0)--(X);
\draw[magenta,line width=0.5mm](X)--(2,0)--(2,-1);
\draw(2,0)--(3,0);
\draw[magenta,line width=0.5mm](1,-1)--(2,-1);
\draw(2,-1)--(3,-1);
\draw[magenta,line width=0.5mm](3,-1)--(4,-1);
\draw[magenta,line width=0.5mm](3,-1)--(3,-2);

\end{tikzpicture} $=$\hspace{1cm}\begin{tikzpicture}[baseline={([yshift=-.5ex]current bounding box.center)}]

\centering

\path (0,0) (1,0) node[circle,inner sep=1pt, draw](X){$X$};

\draw (0,0)--(X);
\draw[magenta,line width=0.5mm] (X)--(2,-0.5);
\draw (3,-0.5)--(4,0)--(5,0);
\draw (2,-0.5)--(3,-0.5);
\draw[magenta,line width=0.5mm] (2,-0.5)--(1,-1)--(0,-1);
\draw (3,-0.5)--(4,-1)--(4.5,-1);
\draw [magenta,line width=0.5mm](5,-1)--(4.5,-1)--(4.5,-2);

\end{tikzpicture} \hspace{1cm} $=$ \begin{tikzpicture}[baseline={([yshift=-.5ex]current bounding box.center)}]

\centering

\path (2,0.5) node[circle,inner sep=1pt, draw](X){$X$};

\draw (0,0) -- (1,0);
\draw(1,0)--(1.5,0.5)--(X);
\draw[magenta,line width=0.5mm] (X)--(2.5,0.5)--(3,0);
\draw(3,0)--(4,0);
\draw(1,0)--(2,-0.5);
\draw[magenta,line width=0.5mm](2,-0.5)--(3,0);
\draw[magenta,line width=0.5mm](2,-0.5)--(2,-1.5);
\end{tikzpicture} \hspace{0.5cm}$= $ \hspace{1cm} \begin{tikzpicture}[baseline={([yshift=-.5ex]current bounding box.center)}]

\centering

\path (1,0) node[circle,inner sep=1pt, draw](X){$X$};
\draw (0,0) -- (X);
\draw[magenta,line width=0.5mm] (X)--(2,0);
\draw[magenta,line width=0.5mm](2,0)--(3,0);
\draw(3,0)--(4,0);
\draw(2,0)--(2.5,-1);
\draw[magenta,line width=0.5mm](2.5,-1)--(3,0);
\draw[magenta,line width=0.5mm] (2.5,-1)--(2.5,-2);

\end{tikzpicture} $=$\hspace{1cm}\begin{tikzpicture}[baseline={([yshift=-.5ex]current bounding box.center)}]

\centering

\path (1,0) node[circle,inner sep=1pt, draw](X){$X$};
\draw(0,0) -- (X);
\draw[magenta,line width=0.5mm] (X)--(3,0)--(3,-1);
\draw(3,0)--(4,0);
\end{tikzpicture}

\vspace{0.5cm} $= \psi_3$
\caption{Diagrammatic proof of $C_4 \langle \psi_3| =  \langle \psi_3|$.}
\label{fig:p6}
\end{figure}

\section{Necessity of supercommutativity}\label{MSproof}

\begin{figure}[H]
\centering

\begin{subfigure}{0.5\textwidth}
\centering

\begin{tikzpicture} 

\draw[decoration={markings, mark=at position 0.5 with {\arrow[scale=2]{>}}}, postaction={decorate}] (0,1) -- (1,1);
\draw[decoration={markings, mark=at position 0.6 with {\arrow[scale=2]{>}}}, postaction={decorate}] (1,1) -- (2,1);
\draw[decoration={markings, mark=at position 0.7 with {\arrow[scale=2]{>}}}, postaction={decorate}] (2,1) -- (5,1);
\draw[decoration={markings, mark=at position 0.8 with {\arrow[scale=2]{>}}}, postaction={decorate}] (5,1) -- (6,1);

\draw[decoration={markings, mark=at position 0.5 with {\arrow[scale=2]{<}}}, postaction={decorate}] (2,1) arc (180:15:0.75);
\draw (5,1) arc (360:195:0.75);

\draw[decoration={markings, mark=at position 0.5 with {\arrow[scale=2]{>}}}, postaction={decorate}] (0,0) arc (270:360:1);

\draw[magenta,line width=1mm] (1,1) arc (360:342:1);
\draw[magenta,line width=1mm] (2,1) -- (1.7,1);
\draw[magenta,line width=1mm] (5,1) arc (360:338:0.75);

\end{tikzpicture}

\vspace{2mm}

\begin{tikzpicture} 

\draw[decoration={markings, mark=at position 0.5 with {\arrow[scale=2]{>}}}, postaction={decorate}] (0,1) -- (1,1);
\draw[decoration={markings, mark=at position 0.6 with {\arrow[scale=2]{>}}}, postaction={decorate}] (1,1) -- (2,1);
\draw[decoration={markings, mark=at position 0.7 with {\arrow[scale=2]{>}}}, postaction={decorate}] (2,1) -- (5,1);
\draw[decoration={markings, mark=at position 0.8 with {\arrow[scale=2]{>}}}, postaction={decorate}] (5,1) -- (6,1);

\draw[decoration={markings, mark=at position 0.5 with {\arrow[scale=2]{<}}}, postaction={decorate}] (2,1) arc (180:15:0.75);
\draw (5,1) arc (360:195:0.75);

\draw[decoration={markings, mark=at position 0.3 with {\arrow[scale=2]{>}}}, postaction={decorate}] (0,0) arc (270:360:1);

\draw[magenta,line width=1mm] (1,1) -- (1.3,1);
\draw[magenta,line width=1mm] (2,1) -- (1.7,1);
\draw[magenta,line width=1mm] (5,1) -- (4.7,1);

\draw [fill] (0.8,0.4) circle (0.1);
\draw [fill] (5.4,1) circle (0.1);
\draw [fill] (4.25,0.25) circle (0.1);

\end{tikzpicture}

\vspace{2mm}

\begin{tikzpicture} 

\draw[decoration={markings, mark=at position 0.25 with {\arrow[scale=2]{>}}}, postaction={decorate}] (0,1) -- (2,1);
\draw[decoration={markings, mark=at position 0.6 with {\arrow[scale=2]{>}}}, postaction={decorate}] (2,1) -- (3,1);
\draw[decoration={markings, mark=at position 0.55 with {\arrow[scale=2]{>}}}, postaction={decorate}] (3,1) -- (5,1);
\draw[decoration={markings, mark=at position 0.8 with {\arrow[scale=2]{>}}}, postaction={decorate}] (5,1) -- (6,1);

\draw[decoration={markings, mark=at position 0.5 with {\arrow[scale=2]{<}}}, postaction={decorate}] (2,1) arc (180:15:0.75);
\draw (5,1) arc (360:195:0.75);

\draw[decoration={markings, mark=at position 0.25 with {\arrow[scale=2]{>}}}, postaction={decorate}] (0,0) -- (2,0);
\draw (2,0) arc (270:360:1);

\draw[magenta,line width=1mm] (3,1) -- (2.7,1);
\draw[magenta,line width=1mm] (2,1) -- (2.3,1);
\draw[magenta,line width=1mm] (5,1) -- (4.7,1);

\draw [fill] (1.3,0) circle (0.1);
\draw [fill] (5.4,1) circle (0.1);

\end{tikzpicture}

\vspace{2mm}

\begin{tikzpicture} 

\draw[decoration={markings, mark=at position 0.25 with {\arrow[scale=2]{>}}}, postaction={decorate}] (0,1) -- (2,1);
\draw[decoration={markings, mark=at position 0.6 with {\arrow[scale=2]{>}}}, postaction={decorate}] (2,1) -- (3,1);
\draw[decoration={markings, mark=at position 0.55 with {\arrow[scale=2]{>}}}, postaction={decorate}] (3,1) -- (5,1);
\draw[decoration={markings, mark=at position 0.8 with {\arrow[scale=2]{>}}}, postaction={decorate}] (5,1) -- (6,1);

\draw[decoration={markings, mark=at position 0.5 with {\arrow[scale=2]{<}}}, postaction={decorate}] (2,1) arc (180:15:0.75);
\draw (5,1) arc (360:195:0.75);

\draw[decoration={markings, mark=at position 0.25 with {\arrow[scale=2]{>}}}, postaction={decorate}] (0,0) -- (2,0);
\draw (2,0) arc (270:360:1);

\draw[magenta,line width=1mm] (3,1) -- (3.3,1);
\draw[magenta,line width=1mm] (2,1) arc (180:158:0.75);
\draw[magenta,line width=1mm] (5,1) -- (4.7,1);

\draw [fill] (5.4,1) circle (0.1);

\end{tikzpicture}

\vspace{2mm}

\begin{tikzpicture} 

\draw[decoration={markings, mark=at position 0.25 with {\arrow[scale=2]{>}}}, postaction={decorate}] (0,1) -- (2,1);
\draw[decoration={markings, mark=at position 0.7 with {\arrow[scale=2]{>}}}, postaction={decorate}] (2,1) -- (5,1);
\draw[decoration={markings, mark=at position 0.8 with {\arrow[scale=2]{>}}}, postaction={decorate}] (5,1) -- (6,1);

\draw[decoration={markings, mark=at position 0.3 with {\arrow[scale=2]{<}}}, postaction={decorate}] (2,1) arc (180:15:0.75);
\draw[decoration={markings, mark=at position 0.85 with {\arrow[scale=2]{>}}}, postaction={decorate}] (5,1) arc (360:195:0.75);

\draw (2.5,2) -- (2,2);
\draw (0,0) arc (270:350:1);
\draw[decoration={markings, mark=at position 0.6 with {\arrow[scale=2]{<}}}, postaction={decorate}] (2,2) arc (90:170:1);
\draw (2.75,1.75) arc (0:90:0.25);

\draw[magenta,line width=1mm] (2,1) arc (180:158:0.75);
\draw[magenta,line width=1mm] (2.75,1.75) arc (90:68:0.75);
\draw[magenta,line width=1mm] (5,1) arc (360:338:0.75);

\end{tikzpicture}

\vspace{2mm}

\begin{tikzpicture} 

\draw[decoration={markings, mark=at position 0.25 with {\arrow[scale=2]{>}}}, postaction={decorate}] (0,1) -- (2,1);
\draw[decoration={markings, mark=at position 0.7 with {\arrow[scale=2]{>}}}, postaction={decorate}] (2,1) -- (5,1);
\draw[decoration={markings, mark=at position 0.8 with {\arrow[scale=2]{>}}}, postaction={decorate}] (5,1) -- (6,1);

\draw[decoration={markings, mark=at position 0.3 with {\arrow[scale=2]{<}}}, postaction={decorate}] (2,1) arc (180:15:0.75);
\draw[decoration={markings, mark=at position 0.85 with {\arrow[scale=2]{>}}}, postaction={decorate}] (5,1) arc (360:195:0.75);

\draw (2.5,2) -- (2,2);
\draw (0,0) arc (270:350:1);
\draw[decoration={markings, mark=at position 0.6 with {\arrow[scale=2]{<}}}, postaction={decorate}] (2,2) arc (90:170:1);
\draw (2.75,1.75) arc (0:90:0.25);

\draw[magenta,line width=1mm] (2,1) arc (180:158:0.75);
\draw[magenta,line width=1mm] (2.75,1.75) arc (90:112:0.75);
\draw[magenta,line width=1mm] (5,1) arc (360:338:0.75);

\draw [fill] (2,2) circle (0.1);
\draw [fill] (4.25,0.25) circle (0.1);

\end{tikzpicture}

\vspace{2mm}

\begin{tikzpicture} 

\draw[decoration={markings, mark=at position 0.5 with {\arrow[scale=2]{>}}}, postaction={decorate}] (0,1) -- (1,1);
\draw[decoration={markings, mark=at position 0.6 with {\arrow[scale=2]{>}}}, postaction={decorate}] (1,1) -- (2,1);
\draw[decoration={markings, mark=at position 0.7 with {\arrow[scale=2]{>}}}, postaction={decorate}] (2,1) -- (5,1);
\draw[decoration={markings, mark=at position 0.8 with {\arrow[scale=2]{>}}}, postaction={decorate}] (5,1) -- (6,1);

\draw[decoration={markings, mark=at position 0.5 with {\arrow[scale=2]{<}}}, postaction={decorate}] (2,1) arc (180:15:0.75);
\draw (5,1) arc (360:195:0.75);

\draw (0.62,0.62) -- (0.62,0.85);
\draw[decoration={markings, mark=at position 0.6 with {\arrow[scale=2]{>}}}, postaction={decorate}] (0,0) arc (270:360:0.62);
\draw (0.62,1.15) arc (180:0:0.19);
\draw (1,1.15) -- (1,1);

\draw[magenta,line width=1mm] (2,1) -- (1.7,1);
\draw[magenta,line width=1mm] (1,1) -- (1.3,1);
\draw[magenta,line width=1mm] (5,1) arc (360:338:0.75);

\draw [fill] (0.62,0.62) circle (0.1);

\end{tikzpicture}

\vspace{2mm}

\begin{tikzpicture} 

\draw[decoration={markings, mark=at position 0.5 with {\arrow[scale=2]{>}}}, postaction={decorate}] (0,1) -- (1,1);
\draw[decoration={markings, mark=at position 0.6 with {\arrow[scale=2]{>}}}, postaction={decorate}] (1,1) -- (2,1);
\draw[decoration={markings, mark=at position 0.7 with {\arrow[scale=2]{>}}}, postaction={decorate}] (2,1) -- (5,1);
\draw[decoration={markings, mark=at position 0.8 with {\arrow[scale=2]{>}}}, postaction={decorate}] (5,1) -- (6,1);

\draw[decoration={markings, mark=at position 0.5 with {\arrow[scale=2]{<}}}, postaction={decorate}] (2,1) arc (180:15:0.75);
\draw (5,1) arc (360:195:0.75);

\draw (0.62,0.62) -- (0.62,0.85);
\draw[decoration={markings, mark=at position 0.6 with {\arrow[scale=2]{>}}}, postaction={decorate}] (0,0) arc (270:360:0.62);
\draw (0.62,1.15) arc (180:0:0.19);
\draw (1,1.15) -- (1,1);

\draw[magenta,line width=1mm] (2,1) -- (1.7,1);
\draw[magenta,line width=1mm] (1,1) -- (0.7,1);
\draw[magenta,line width=1mm] (5,1) arc (360:338:0.75);

\draw [fill] (0.2,1) circle (0.1);

\end{tikzpicture}

\caption{NS sector: $C_{ij}=(-1)^{|i||j|+|i|}C_{ji}$}
\end{subfigure}%
\begin{subfigure}{0.5\textwidth}
\centering

\begin{tikzpicture} 

\draw[decoration={markings, mark=at position 0.5 with {\arrow[scale=2]{>}}}, postaction={decorate}] (0,1) -- (1,1);
\draw[decoration={markings, mark=at position 0.6 with {\arrow[scale=2]{>}}}, postaction={decorate}] (1,1) -- (2,1);
\draw[decoration={markings, mark=at position 0.7 with {\arrow[scale=2]{>}}}, postaction={decorate}] (2,1) -- (5,1);
\draw[decoration={markings, mark=at position 0.8 with {\arrow[scale=2]{>}}}, postaction={decorate}] (5,1) -- (6,1);

\draw[decoration={markings, mark=at position 0.5 with {\arrow[scale=2]{<}}}, postaction={decorate}] (2,1) arc (180:15:0.75);
\draw (5,1) arc (360:195:0.75);

\draw[decoration={markings, mark=at position 0.5 with {\arrow[scale=2]{>}}}, postaction={decorate}] (0,0) arc (270:360:1);

\draw[magenta,line width=1mm] (1,1) arc (360:342:1);
\draw[magenta,line width=1mm] (2,1) -- (1.7,1);
\draw[magenta,line width=1mm] (5,1) arc (360:338:0.75);

\draw [fill] (4.25,0.25) circle (0.1);

\end{tikzpicture}

\vspace{2mm}

\begin{tikzpicture} 

\draw[decoration={markings, mark=at position 0.5 with {\arrow[scale=2]{>}}}, postaction={decorate}] (0,1) -- (1,1);
\draw[decoration={markings, mark=at position 0.6 with {\arrow[scale=2]{>}}}, postaction={decorate}] (1,1) -- (2,1);
\draw[decoration={markings, mark=at position 0.7 with {\arrow[scale=2]{>}}}, postaction={decorate}] (2,1) -- (5,1);
\draw[decoration={markings, mark=at position 0.8 with {\arrow[scale=2]{>}}}, postaction={decorate}] (5,1) -- (6,1);

\draw[decoration={markings, mark=at position 0.5 with {\arrow[scale=2]{<}}}, postaction={decorate}] (2,1) arc (180:15:0.75);
\draw (5,1) arc (360:195:0.75);

\draw[decoration={markings, mark=at position 0.3 with {\arrow[scale=2]{>}}}, postaction={decorate}] (0,0) arc (270:360:1);

\draw[magenta,line width=1mm] (1,1) -- (1.3,1);
\draw[magenta,line width=1mm] (2,1) -- (1.7,1);
\draw[magenta,line width=1mm] (5,1) -- (4.7,1);

\draw [fill] (0.8,0.4) circle (0.1);
\draw [fill] (5.4,1) circle (0.1);

\end{tikzpicture}

\vspace{2mm}

\begin{tikzpicture} 

\draw[decoration={markings, mark=at position 0.25 with {\arrow[scale=2]{>}}}, postaction={decorate}] (0,1) -- (2,1);
\draw[decoration={markings, mark=at position 0.6 with {\arrow[scale=2]{>}}}, postaction={decorate}] (2,1) -- (3,1);
\draw[decoration={markings, mark=at position 0.55 with {\arrow[scale=2]{>}}}, postaction={decorate}] (3,1) -- (5,1);
\draw[decoration={markings, mark=at position 0.8 with {\arrow[scale=2]{>}}}, postaction={decorate}] (5,1) -- (6,1);

\draw[decoration={markings, mark=at position 0.5 with {\arrow[scale=2]{<}}}, postaction={decorate}] (2,1) arc (180:15:0.75);
\draw (5,1) arc (360:195:0.75);

\draw[decoration={markings, mark=at position 0.25 with {\arrow[scale=2]{>}}}, postaction={decorate}] (0,0) -- (2,0);
\draw (2,0) arc (270:360:1);

\draw[magenta,line width=1mm] (3,1) -- (2.7,1);
\draw[magenta,line width=1mm] (2,1) -- (2.3,1);
\draw[magenta,line width=1mm] (5,1) -- (4.7,1);

\draw [fill] (1.3,0) circle (0.1);
\draw [fill] (5.4,1) circle (0.1);
\draw [fill] (4.25,0.25) circle (0.1);

\end{tikzpicture}

\vspace{2mm}

\begin{tikzpicture} 

\draw[decoration={markings, mark=at position 0.25 with {\arrow[scale=2]{>}}}, postaction={decorate}] (0,1) -- (2,1);
\draw[decoration={markings, mark=at position 0.6 with {\arrow[scale=2]{>}}}, postaction={decorate}] (2,1) -- (3,1);
\draw[decoration={markings, mark=at position 0.55 with {\arrow[scale=2]{>}}}, postaction={decorate}] (3,1) -- (5,1);
\draw[decoration={markings, mark=at position 0.8 with {\arrow[scale=2]{>}}}, postaction={decorate}] (5,1) -- (6,1);

\draw[decoration={markings, mark=at position 0.5 with {\arrow[scale=2]{<}}}, postaction={decorate}] (2,1) arc (180:15:0.75);
\draw (5,1) arc (360:195:0.75);

\draw[decoration={markings, mark=at position 0.25 with {\arrow[scale=2]{>}}}, postaction={decorate}] (0,0) -- (2,0);
\draw (2,0) arc (270:360:1);

\draw[magenta,line width=1mm] (3,1) -- (3.3,1);
\draw[magenta,line width=1mm] (2,1) arc (180:158:0.75);
\draw[magenta,line width=1mm] (5,1) -- (4.7,1);

\draw [fill] (5.4,1) circle (0.1);
\draw [fill] (4.25,0.25) circle (0.1);

\end{tikzpicture}

\vspace{2mm}

\begin{tikzpicture} 

\draw[decoration={markings, mark=at position 0.25 with {\arrow[scale=2]{>}}}, postaction={decorate}] (0,1) -- (2,1);
\draw[decoration={markings, mark=at position 0.7 with {\arrow[scale=2]{>}}}, postaction={decorate}] (2,1) -- (5,1);
\draw[decoration={markings, mark=at position 0.8 with {\arrow[scale=2]{>}}}, postaction={decorate}] (5,1) -- (6,1);

\draw[decoration={markings, mark=at position 0.3 with {\arrow[scale=2]{<}}}, postaction={decorate}] (2,1) arc (180:15:0.75);
\draw[decoration={markings, mark=at position 0.85 with {\arrow[scale=2]{>}}}, postaction={decorate}] (5,1) arc (360:195:0.75);

\draw (2.5,2) -- (2,2);
\draw (0,0) arc (270:350:1);
\draw[decoration={markings, mark=at position 0.6 with {\arrow[scale=2]{<}}}, postaction={decorate}] (2,2) arc (90:170:1);
\draw (2.75,1.75) arc (0:90:0.25);

\draw[magenta,line width=1mm] (2,1) arc (180:158:0.75);
\draw[magenta,line width=1mm] (2.75,1.75) arc (90:68:0.75);
\draw[magenta,line width=1mm] (5,1) arc (360:338:0.75);

\draw [fill] (4.25,0.25) circle (0.1);
\draw [fill] (2,2) circle (0.1);

\end{tikzpicture}

\vspace{2mm}

\begin{tikzpicture} 

\draw[decoration={markings, mark=at position 0.25 with {\arrow[scale=2]{>}}}, postaction={decorate}] (0,1) -- (2,1);
\draw[decoration={markings, mark=at position 0.7 with {\arrow[scale=2]{>}}}, postaction={decorate}] (2,1) -- (5,1);
\draw[decoration={markings, mark=at position 0.8 with {\arrow[scale=2]{>}}}, postaction={decorate}] (5,1) -- (6,1);

\draw[decoration={markings, mark=at position 0.3 with {\arrow[scale=2]{<}}}, postaction={decorate}] (2,1) arc (180:15:0.75);
\draw[decoration={markings, mark=at position 0.85 with {\arrow[scale=2]{>}}}, postaction={decorate}] (5,1) arc (360:195:0.75);

\draw (2.5,2) -- (2,2);
\draw (0,0) arc (270:350:1);
\draw[decoration={markings, mark=at position 0.6 with {\arrow[scale=2]{<}}}, postaction={decorate}] (2,2) arc (90:170:1);
\draw (2.75,1.75) arc (0:90:0.25);

\draw[magenta,line width=1mm] (2,1) arc (180:158:0.75);
\draw[magenta,line width=1mm] (2.75,1.75) arc (90:112:0.75);
\draw[magenta,line width=1mm] (5,1) arc (360:338:0.75);

\end{tikzpicture}

\vspace{2mm}

\begin{tikzpicture} 

\draw[decoration={markings, mark=at position 0.5 with {\arrow[scale=2]{>}}}, postaction={decorate}] (0,1) -- (1,1);
\draw[decoration={markings, mark=at position 0.6 with {\arrow[scale=2]{>}}}, postaction={decorate}] (1,1) -- (2,1);
\draw[decoration={markings, mark=at position 0.7 with {\arrow[scale=2]{>}}}, postaction={decorate}] (2,1) -- (5,1);
\draw[decoration={markings, mark=at position 0.8 with {\arrow[scale=2]{>}}}, postaction={decorate}] (5,1) -- (6,1);

\draw[decoration={markings, mark=at position 0.5 with {\arrow[scale=2]{<}}}, postaction={decorate}] (2,1) arc (180:15:0.75);
\draw (5,1) arc (360:195:0.75);

\draw (0.62,0.62) -- (0.62,0.85);
\draw[decoration={markings, mark=at position 0.6 with {\arrow[scale=2]{>}}}, postaction={decorate}] (0,0) arc (270:360:0.62);
\draw (0.62,1.15) arc (180:0:0.19);
\draw (1,1.15) -- (1,1);

\draw[magenta,line width=1mm] (2,1) -- (1.7,1);
\draw[magenta,line width=1mm] (1,1) -- (1.3,1);
\draw[magenta,line width=1mm] (5,1) arc (360:338:0.75);

\draw [fill] (4.25,0.25) circle (0.1);

\end{tikzpicture}

\vspace{2mm}

\begin{tikzpicture} 

\draw[decoration={markings, mark=at position 0.5 with {\arrow[scale=2]{>}}}, postaction={decorate}] (0,1) -- (1,1);
\draw[decoration={markings, mark=at position 0.6 with {\arrow[scale=2]{>}}}, postaction={decorate}] (1,1) -- (2,1);
\draw[decoration={markings, mark=at position 0.7 with {\arrow[scale=2]{>}}}, postaction={decorate}] (2,1) -- (5,1);
\draw[decoration={markings, mark=at position 0.8 with {\arrow[scale=2]{>}}}, postaction={decorate}] (5,1) -- (6,1);

\draw[decoration={markings, mark=at position 0.5 with {\arrow[scale=2]{<}}}, postaction={decorate}] (2,1) arc (180:15:0.75);
\draw (5,1) arc (360:195:0.75);

\draw (0.62,0.62) -- (0.62,0.85);
\draw[decoration={markings, mark=at position 0.6 with {\arrow[scale=2]{>}}}, postaction={decorate}] (0,0) arc (270:360:0.62);
\draw (0.62,1.15) arc (180:0:0.19);
\draw (1,1.15) -- (1,1);

\draw[magenta,line width=1mm] (2,1) -- (1.7,1);
\draw[magenta,line width=1mm] (1,1) -- (0.7,1);
\draw[magenta,line width=1mm] (5,1) arc (360:338:0.75);

\draw [fill] (4.25,0.25) circle (0.1);

\end{tikzpicture}

\caption{R sector: $C_{ij}=(-1)^{|i||j|}C_{ji}$}
\end{subfigure}%

\caption{A proof of equations \eqref{ms1} and \eqref{ms2}. Arrows denote edge directions, magenta line segments denote special edges, and black dots denote spin signs $+1$, i.e. insertions of $\cF$.}
\label{fig:msproof}
\end{figure}
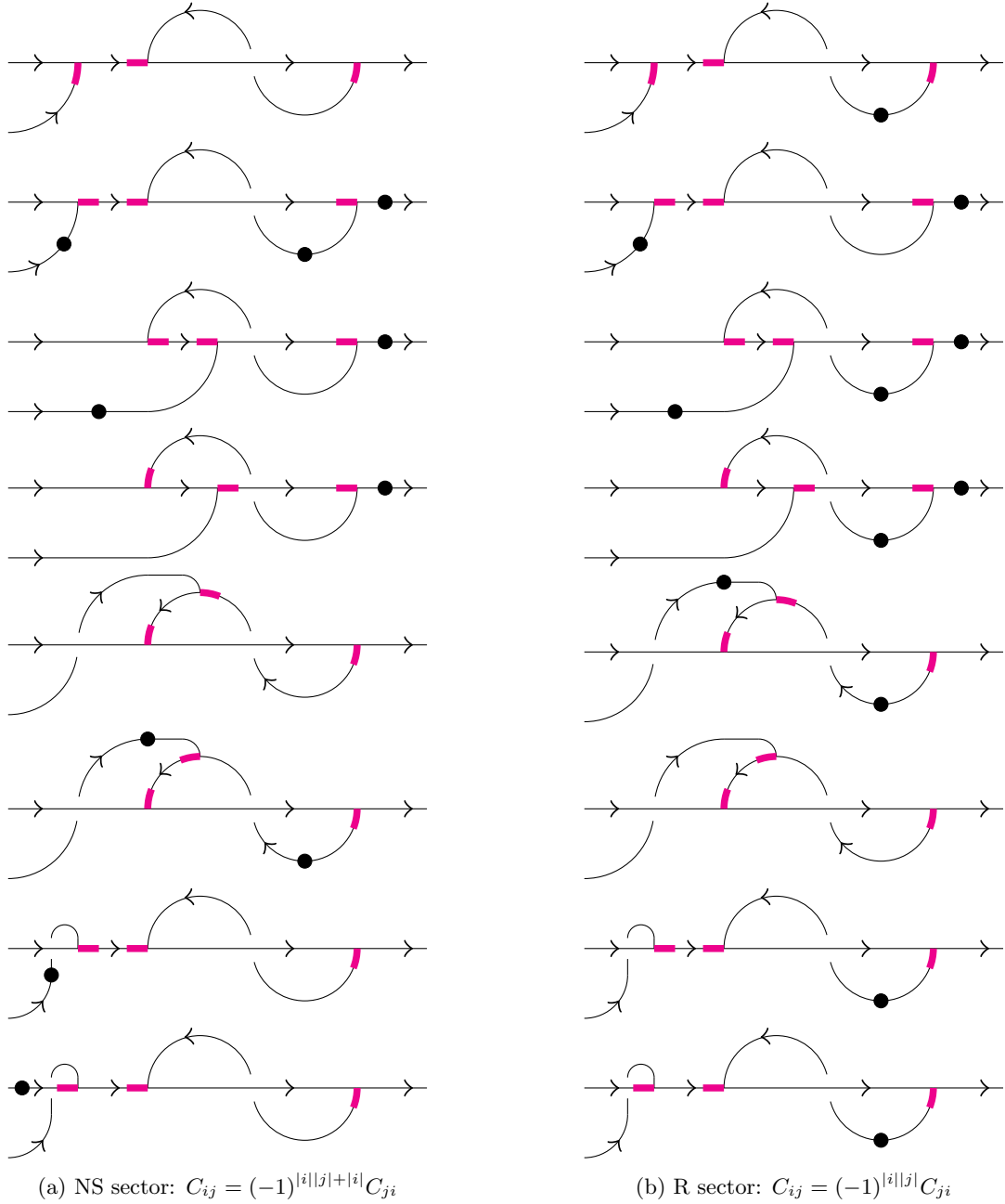

This appendix is a derivation the results \eqref{ms1} and \eqref{ms2} from the lattice spin formalism introduced in Section \ref{spintqft}. Consider acting on the state $\ket{ij}$ with the cylinder map $Z(C)$; this is represented in the top diagram of each column of Figure \ref{fig:msproof}. To manipulate these diagrams into the diagrams at the bottom of each column, one applies a series of ``moves'' that are like Pachner moves but are compatible with the lattice spin structure (see \cite{NR} for details). Finally, one unbraids the legs at the cost of a sign $(-1)^{|i||j|}$.

\section{Description of $\omega$ in terms of pairs $(\alpha, \beta)$}\label{isom}


Start with some $[\omega] \in H^2(\mathcal{G}, U(1))$. We denote by $\bar{g}$ either an element of $G_b$ or the corresponding element in $\mathcal{G}$ whose $t(g) = 0$, i.e. $(\bar{g},0)$. A general element of $\mathcal{G}$ then takes the form of either $\bar{g}$ or $\bar{g} p$. 

Given an arbitrary $\omega$, we can shift it by a coboundary $\delta B$ where  $ B \in C^1(\mathbb{Z}_2, U(1))$ such that $B(0) = 0$ and $B(p) = \frac12  \omega(p,p)$ so that our new $\omega$ satisfies $\omega(p,p)=0$. Then we can add a coboundary $\delta A$ with $A \in C^1(\mathcal{G}, \mathbb{Z}_2)$ satisfying $A(\bar{g}p) = A(\bar{g}) - \omega(\bar{g},p)$ to $\omega$ to make $\omega(\bar{g},p)=0$ for all $\bar{g} \in G_b$. 

Evaluating the $3$-cochain $\delta \omega$ on $(\bar{g},p,p)$, $(\bar{g},\bar{h},p)$, and $(\bar{g}p, \bar{h}, p)$, and using the fact that $\delta \omega = 0$, we see that changing the second argument of $\omega$ by $p$ does not affect its value, i.e. $\omega(g, h) = \omega(g,hp)$, $\forall g,h \in \mathcal{G}$.

Then, evaluating $\delta \omega$ on $(\bar{g}, p, \bar{h})$ gives $\omega(\bar{g}p,\bar{h}) = \omega(\bar{g},\bar{h})+ \omega(p,\bar{h})$. Defining $\alpha (\bar{g}, \bar{h}) :=  \omega(\bar{g},\bar{h})$ and $\beta(\bar{g}) := \omega(p, \bar{g})$, $\omega = \alpha + t \cup \beta$, and we can check that $\delta \beta =0$ and hence $\delta \alpha = -\delta t \cup \beta = \rho \cup \beta$. With our gauge choice, one can show that this definition of $\beta$ is consistent with $\beta(\bar{g}) = |Q(g)|$. The residual gauge freedom which shifts $\omega$ by a coboundary $\delta \lambda$ for $\lambda$ which is a pull-back from $G_b$. This leaves $\beta$ invariant but shifts $\alpha$ by a $G_b$-coboundary. Hence $\alpha \sim \alpha + \delta \lambda$, and we see that equivalence classes of $\omega$ correspond to equivalence classes of pairs $(\alpha, \beta)$ satisfying $\delta \alpha =  \rho \cup \beta$ and $\delta \beta =0$ with $(\alpha, \beta) \sim (\alpha + \delta \lambda, \beta)$.

When $\mathcal{G}$ splits, $\rho$ is trivial and we have $\delta \alpha =0$, so the set of equivalence classes of $\alpha$ is $H^2(G, U(1))$. The set of equivalence classes of $\beta$ is of course $H^1(G_b, \mathbb{Z}_2)$. This confirms $H^2(\mathcal{G}, U(1)) \simeq H^2(G_b, U(1)) \times H^1(G_b, \mathbb{Z}_2)$, which we already knew from more abstract arguments.

\section{Derivation of the group law for fermionic SRE phases}\label{sec:grouplaw}

In the body of the paper we derived the supertensor product of two $\cG$-graded algebras of the form $\End(U_i)$, $i=1,2$, where $(Q_i,U_i)$ is a projective representation of $\cG=G_b\times\ZZ_2$. This allowed us to determine the group law for $\gamma=0$ SRE phases. Here we compute the supertensor product for $\cG$-equivariant algebras involving a $\Cl(1)$ factor and determine the group law in the remaining cases.

Let $(Q_1,U_1)$ be a projective representation of $\cG$ with a 2-cocycle parameterized by a pair $(\alpha_1,\beta_1)\in Z^2(G_b,U(1))\times Z^1(G_b,\ZZ_2)$. We will denote $Q_1(p)=P$, so that
\begin{equation}
Q_1(g)Q_1(h)=\exp(2\pi i \alpha_1(g,h)) Q_1(gh),\quad P Q_1(g) P^{-1}=(-1)^{\beta_1(\bar g)}Q_1(g).
\end{equation}
Let $(Q_2,U_2)$ be a projective representation of $G_b$ with a 2-cocycle $\alpha_2\in Z^2(G_b,U(1))$, i.e.
\begin{equation}
Q_2(g)Q_2(h)=\exp(2\pi i \alpha_2(g,h)) Q_2(gh)
\end{equation}
The vector space $U_2$ is regarded as purely even. Let $\beta_2:G_b\ra\ZZ_2$ be a homomorphism. Let $A_1$ be the algebra $\End(U_1)$ with the obvious $\cG$ action. Let $A_2=\End(U_2)\otimes \Cl(1)$, and define a $\cG$ action on it as follows:
\begin{equation}
g: M\otimes \Gamma^m\mapsto (-1)^{m \beta_2(g)} Q_2(g) M Q_2(g)^{-1}\otimes \Gamma^m,
\end{equation}
and 
\begin{equation}
p: M\otimes \Gamma^m\mapsto (-1)^m M\otimes\Gamma^m.
\end{equation}
where $M\in \End(U_2), m\in\ZZ_2$.

The first claim is that $A_1\fotimes A_2$ is isomorphic (as a $\ZZ_2$-graded algebra) to $A_{12}=\End(U_1\otimes U_2)\otimes \Cl(1)$, where both $U_1$ and $U_2$ are regarded as purely even. The isomorphism is given by
\begin{equation}
JW: M_1\fotimes M_2\fotimes\Gamma^m\mapsto M_1 P^m\otimes M_2\otimes\Gamma^{m+|M_1|} 
\end{equation}
We denoted it $JW$ to indicate that it is a version of the Jordan-Wigner transformation. It is easy to check that the map preserves the product as well as grading, and its inverse is
\begin{equation}
JW^{-1}: M_1\otimes M_2\otimes\Gamma^m\mapsto  M_1 P^{m+|M_1|}\fotimes M_2\fotimes\Gamma^{m+|M_1|}
\end{equation}
Thus the parameter $\gamma$ for $A_{12}$ is $1$.

Next we compute the action of $G_b$ on $A_{12}$ induced by the isomorphism $JW$. We get:
\begin{equation}
JW\circ g\circ JW^{-1}: M_1\otimes M_2\otimes\Gamma^m\mapsto (-1)^{(\beta_1(\bar g)+\beta_2(\bar g))(m+|M_1|)} Q_1(g) M_1 Q_1(g)^{-1}\otimes Q_2(g) M_2 Q_2(g)^{-1}\otimes\Gamma^m.
\end{equation}
To bring this $G_b$-action to the standard form, we define ${\tilde Q}_1(g)=Q_1(g) P^{\beta_1(g)+\beta_2(g)} i^{\beta_1(g)}$. Then the $G_b$-action on $\End(U)\otimes\Cl(1)$ takes the form
\begin{equation}
M_1\otimes M_2\otimes \Gamma^m\mapsto (-1)^{m(\beta_1(\bar g)+\beta_2(\bar g))} {\tilde Q}_1(g) M_1 {\tilde Q_1}(g)^{-1} \otimes Q_2(g) M_2 Q_2(g)^{-1}\otimes\Gamma^m.
\end{equation}
Thus the parameter $\beta$ for $A_{12}$ is $\beta_1+\beta_2$. Finally, it is easy to check that the matrices ${\tilde Q}_1(g)\otimes Q_2(g)$ form a projective representation of $G_b$ with a 2-cocycle
\begin{equation}
\alpha(g,h)=\alpha_1(g,h)+\alpha_2(g,h)+\frac12\beta_1(h)\beta_2(g).
\end{equation}
We conclude that the group law for the parameters $(\alpha,\beta,\gamma)$ obeys
\begin{equation}
(\alpha_1,\beta_1,0)+(\alpha_2,\beta_2,1)=(\alpha_1+\alpha_2+\frac12 \beta_1\cup\beta_2,\beta_1+\beta_2,1).
\end{equation}

The last case to consider is $\gamma_1=\gamma_2=1$. The algebrs to be tensored are $A_1=\End(U_1)\otimes\Cl(1)$ and $A_2=\End(U_2)\otimes\Cl(1)$, where $(Q_1,U_1)$ and $(Q_2,U_2)$ are projective representations of $G_b$ with 2-cocycles $\alpha_1$ and $\alpha_2$. The group $G_b$ acts as follows on the generators of the two Clifford algebras:
\begin{equation}\label{cliffaction}
g: \Gamma_i\mapsto (-1)^{\beta_i(\bar g)}\Gamma_i,\quad i=1,2. 
\end{equation}
It is easy to see that $\Cl(1)\fotimes\Cl(1)=\Cl(2)$, and that $\Cl(2)\simeq \End(\CC^2)$. The isomorphism sends $\Gamma_i$ to $\sigma_i$, $i=1,2$, and the action of $p$ on $\CC^2$ is given by the Pauli matrix $\sigma_3=-i\Gamma_1\Gamma_2$. Thus 
\begin{equation}
A_{12}=A_1\fotimes A_2\simeq \End(U_1\otimes U_2\otimes \CC^2),
\end{equation}
where $U_1$ and $U_2$ are regarded as purely even. Thus the $\gamma$ parameter for $A_{12}$ is $0$. 

The group $G_b$ acts on $U_1\otimes U_2$ by $Q_1\otimes Q_2$. This is a projective action, with a 2-cocycle $\alpha_1+\alpha_2$. There is no canonical choice of the projective $G_b$ action on $\CC^2$ which induces the action (\ref{cliffaction}) on $\Cl(2)\simeq \End(\CC^2)$. One possible choice is
\begin{equation}\label{possibleaction}
g: v\mapsto \Gamma_1^{\beta_2(\bar g)}\Gamma_2^{\beta_1(\bar g)} v,\quad v\in\CC^2.
\end{equation}
Any other choice differs from this one by a scalar factor $\exp(\lambda(g))$ which changes the corresponding 2-cocycle by a coboundary. Using the action (\ref{possibleaction}), the corresponding 2-cocycle is $\frac12\beta_1(\bar g)\beta_2(\bar h)$. The net result is that the $G_b$ action on $U_1\otimes U_2\otimes \CC^2$ is projective with a 2-cocycle 
$\alpha_1+\alpha_2+\frac12 \beta_1\cup\beta_2$. We also compute:
\begin{equation}
(-i\Gamma_1\Gamma_2)\Gamma_1^{\beta_2(\bar g)}\Gamma_2^{\beta_1(\bar g)}=(-1)^{\beta_1(\bar g)+\beta_2(\bar g)}\Gamma_1^{\beta_2(\bar g)}\Gamma_2^{\beta_1(\bar g)} (-i\Gamma_1\Gamma_2).
\end{equation}
This implies that the parameter $\beta$ for $A_{12}$ is $\beta_1+\beta_2$.

We have shown that for the special case $\gamma_1=\gamma_2=1$ the group law 
says
\begin{equation}
(\alpha_1,\beta_1,1)+(\alpha_2,\beta_2,1)=(\alpha_1+\alpha_2+\frac12\beta_1\cup\beta_2,\beta_1+\beta_2,0)
\end{equation}
This completes the proof of (\ref{groupstructure}).

\end{document}